\documentstyle[pra,aps,manuscript]{revtex}

\input{epsf}

\begin{document}

\draft 

\title{Entangled-Photon Generation from Parametric Down-Conversion in Media with
Inhomogeneous Nonlinearity\\}

\author{Giovanni~Di~Giuseppe,$^1$\thanks{Also with Istituto Elettrotecnico Nazionale
{\it G.~Ferraris}, Strada delle Cacce 91, I-10153 Torino, Italy.}
Mete~Atat\"{u}re,$^2$ Matthew~D.~ Shaw,$^1$
Alexander~V.~Sergienko,$^{1,2}$ Bahaa~E.~A.~Saleh,$^1$ and
Malvin~C.~Teich$^{1,2}$}

\address {Quantum Imaging Laboratory,\thanks{http://www.bu.edu/qil}\\ $^1$Department of Electrical \& Computer Engineering and $^2$Department of Physics ,\\ Boston
University, 8 Saint Mary's Street, Boston, MA 02215}

\date{\today}

\maketitle

\begin{abstract}

We develop and experimentally verify a theory of Type-II
spontaneous parametric down-conversion (SPDC) in media with
inhomogeneous distributions of second-order nonlinearity. As a
special case, we explore interference effects from SPDC generated
in a cascade of two bulk crystals separated by an air gap. The
polarization quantum-interference pattern is found to vary
strongly with the spacing between the two crystals. This is found
to be a cooperative effect due to two mechanisms: the chromatic
dispersion of the medium separating the crystals and
spatiotemporal effects which arise from the inclusion of
transverse wave vectors. These effects provide two concomitant
avenues for controlling the quantum state generated in SPDC. We
expect these results to be of interest for the development of
quantum technologies and the generation of SPDC in periodically
varying nonlinear materials.

\end{abstract}

\pacs{PACS number(s): {\tt \string 42.50.Dv, 03.65.Bz, 42.65.Ky}}

\narrowtext

\preprint{HEP/123-qed} \vskip1pc

\tableofcontents

\section{Introduction}

Spontaneous parametric down-conversion (SPDC) \cite{SPDC} has now
come into widespread use as a simple and robust source of
entangled photon pairs. Uses for these pairs range from the
examination of quantum mechanical foundations
\cite{Bell,Hardy,GHZ}, to applications in optical measurements
\cite{Metrology}, spectroscopy \cite{Fei-PRL}, imaging
\cite{optical-imaging}, and quantum information
\cite{Cryptography,Teleportation}. As such, there has been
considerable interest in greater optimization and control of the
exotic two-photon states available from SPDC, particularly when
pumped by ultrafast pulses \cite{Fs-SPDC,QIL-fs-SPDC,ata01a}.
Additionally, much work has recently been focused on the use of
cascaded nonlinear
crystals~\cite{mandel,bur97,zeilinger,grice,kwiat,ata01} to
manipulate and improve the generation of the two-photon state.

The photon pairs from Type-II SPDC are generated in a quantum
state which can be entangled in frequency, wave vector, and
polarization. In recent works~\cite{ata01a}, we have demonstrated
the utility of a model which considers entanglement in these
parameters concurrently. It was shown that quantum-interference
patterns were altered predictably by controlling the range of
transverse wave vectors selected by the optical system. In this
paper, we extend this formalism and investigate interference from
SPDC generated in media with inhomogeneous longitudinal
distributions of nonlinearity. The state function of the photon
pair generated in SPDC is completely characterized by three
functions: the spectral profile of the pump, the longitudinal
distribution of nonlinear susceptibility, and the dispersion in
the generation medium. In principle, one could arbitrarily weight
the spatiotemporal distribution of signal and idler modes by a
judicious choice of these three functions. This, in return,
introduces new avenues of control. To demonstrate this, we
consider a rudimentary case of an inhomogeneous medium, two bulk
crystals separated by a linear medium such as an air gap. In this
configuration, a host of interesting effects emerge, such as the
modulation of interference visibility with crystal separation.
This effect and others are theoretically predicted and
experimentally verified in this paper.

The promise of a source whose degree of entanglement is
controllable in frequencies and wave vectors by turning a single
knob is clearly alluring for purposes of quantum-information
processing. The results reported in this paper are also likely to
be of use in guiding future developments in quantum-state
synthesis involving multi-crystal configurations
\cite{mandel,bur97,zeilinger,grice,kwiat,ata01,synthesis}, in
ultrafast-pumped parametric down-conversion
\cite{Fs-SPDC,QIL-fs-SPDC}, and periodically poled materials
\cite{fejer-PPLN,gisin-PPLN}.

\section{Theory}

Our theory considers a quantum state which can be concurrently
entangled in polarization, frequency, and transverse wave vector,
so as to be valid for an arbitrary optical system. As we shall
see, the longitudinal distribution of nonlinearity provides a
powerful means for controlling the structure of the two-photon
quantum state generated in SPDC. As an important special case, we
consider the simple example of SPDC generation in a cascade of two
bulk crystals separated by a linear medium. We then describe the
quantum interference between the two photons of the SPDC pair as
they propagate through an arbitrary linear optical system. This
formalism allows the quantum interference to be analyzed in the
absence of spectral filters and reduces to the conventionally
established single-mode theory in the small-aperture limit, unless
very thick crystals are used.

\subsection{State Generation in Inhomogeneous Media}\label{StateGenerationInHeterogeneousMedia}

For the sake of simplicity, we consider media where effects from
third- and higher-order susceptibilities are weak and can be
neglected. By virtue of the relatively weak interaction in the
nonlinear crystal, we consider the two-photon state generated
within the confines of first-order time-dependent perturbation
theory. The two-photon state at the output of the nonlinear medium
is found in the interaction picture to be \cite{ata01a}

\begin{equation}\label{Psi-General}
    |\Psi^{(2)}\rangle \propto \int d{\bf q}_{\rm o}d{\bf q}_{\rm e}\,d\omega_{\rm o}d\omega_{\rm e}
        ~\Phi({\bf q}_{\rm o},{\bf q}_{\rm e};\omega_{\rm o},\omega_{\rm e})\
            \hat a^{\dagger}_{\rm o}({\bf q}_{\rm o},\omega_{\rm o})\
            \hat a^{\dagger}_{\rm e}({\bf q}_{\rm e},\omega_{\rm e})\
            |0\rangle\,,
\end{equation}

\noindent where the state function

\begin{equation}\label{Phi-Z-Integral}
    \Phi({\bf q}_{\rm o},{\bf q}_{\rm e};\omega_{\rm o},\omega_{\rm e})
        ~=~{\tilde E}_{\rm p}({\bf q}_{\rm o}+{\bf
        q}_{\rm e};\omega_{\rm o}+\omega_{\rm e})\,
            \int dz\ \chi^{(2)}(z)\
                  e^{i\Delta({\bf q}_{\rm o},{\bf
                  q}_{\rm e};\omega_{\rm o},\omega_{\rm e})z}\,.
\end{equation}

\noindent Here $\chi^{(2)}(z)$ is the distribution of second-order
nonlinearity along the longitudinal axis, ${\tilde E}_{\rm p}({\bf
q}_{\rm p}; \omega_{\rm p})$ is the complex-amplitude profile of
the pump field, ${\bf q}_{j}$ ($j={\rm p,o,e}$) is the transverse
component of the wave vector ${\bf k}_j$ in the medium, and
$\Delta$ is the wave vector mismatch function

\begin{eqnarray}\label{Delta}
  \Delta({\bf q}_{\rm o},{\bf q}_{\rm e};\omega_{\rm o},\omega_{\rm e})=
             \kappa_{\rm p}(\omega_{\rm o}+\omega_{\rm e},{\mathbf q}_{\rm o}+{\mathbf q}_{\rm e})-
             \kappa_{\rm o}\left(\omega_{\rm o},{\mathbf q}_{\rm o}\right)-
             \kappa_{\rm e}\left(\omega_{\rm e},{\mathbf q}_{\rm e}\right),
\end{eqnarray}

\noindent which depends on the dispersiveness of the medium. In
this equation, the longitudinal projections $\kappa_{j}$ ($j={\rm
p,o,e}$) are related to the indices $({\bf q}_j,\omega_j)$ via
\begin{equation}\label{kz-int}
  \kappa_j (\omega_j, {\mathbf q}_j) =
  \sqrt{k_j^2(\omega_j, {\mathbf q}_j) - |{\mathbf q}_j|^2 },
\end{equation}

\noindent where $\omega_{\rm p} = \omega_{\rm o} + \omega_{\rm e}$
and ${\mathbf q}_{\rm p} = {\mathbf q}_{\rm o} + {\mathbf q}_{\rm
e}$. Here the wavenumber $k_j\equiv|{\bf
k}_j|=n[\omega_j,\theta({\bf q}_j)]\,\omega_j / c$, where $c$ is
the speed of light in vacuum, $\theta$ is the angle between ${\bf
k}_{\rm p}$ and the optical axis of the nonlinear crystal, and
$n(\omega_j,\theta)$ is the index of refraction in the nonlinear
medium. Note that the symbol $n(\omega,\theta)$ represents the
extraordinary refractive index $n_{\rm e}(\omega,\theta)$ when
calculating $\kappa$ for extraordinary waves, and the ordinary
refractive index $n_{\rm o}(\omega)$ for ordinary waves.

Note from Eq.~(\ref{Phi-Z-Integral}) that the state function is
completely characterized by the spectral profile of the pump, the
longitudinal distribution of nonlinear susceptibility, and the
dispersion in the generation medium. All three of these parameters
may be controlled experimentally, and all three present avenues
for controlling the structure of the two-photon quantum state.

For a medium with an inhomogeneous distribution of nonlinear
susceptibility along the longitudinal axis, it is convenient to
define

\begin{equation}\label{Chi-Arbitrary}
    \chi^{(2)}(z) = \int d\zeta\,{\tilde
\chi}^{(2)}(\zeta)\,
    e^{-i\zeta\,z}\,,
\end{equation}

\noindent where ${\tilde \chi}^{(2)}(\zeta)$ is the inverse
Fourier transform of $\chi^{(2)}(z)$. Substitution into
Eq.~(\ref{Phi-Z-Integral}) then gives

\begin{equation}\label{FT-Phi-General}
    \Phi({\bf q}_{\rm o},{\bf q}_{\rm e};\omega_{\rm o},\omega_{\rm e})
            ~=~{\tilde E}_{\rm p}({\bf q}_{\rm o}+{\bf q}_{\rm e};\omega_{\rm o}+\omega_{\rm e})\
            {\tilde \chi}^{(2)}[\Delta({\bf q}_{\rm o},{\bf
q}_{\rm e};\omega_{\rm o},\omega_{\rm e})]
\end{equation}

\noindent for a uniformly dispersive medium. For example, a single
bulk crystal of thickness $L$ and constant nonlinearity $\chi_0$
has a nonlinear susceptibility profile $\chi^{(2)}(z) = \chi_{0}
{\rm rect}_{[-L,0]}(z)$ where ${\rm rect}_{[-L,0]}(z)=1$ if $-L
\leq z \leq 0$ and zero otherwise. In this case, the inverse
Fourier transform of the nonlinearity profile becomes

\begin{equation}\label{Chi-General}
    {\tilde \chi}^{(2)}(\Delta)
            = \chi_0 L\ {\rm sinc}\left(\frac{L\Delta}{2}\right)
                e^{{\rm -i}\frac{L\Delta}{2}}\,.
\end{equation}

\noindent For a monochromatic plane wave pump with a central
frequency $\omega_{\rm p}^{0}$, ${\tilde E}_{\rm p}({\bf q}_{\rm
p}; \omega_{\rm p})$ in Eq.~(\ref{FT-Phi-General}) is proportional
to $\delta({\bf q}_{\rm o}+{\bf q}_{\rm e})\delta(\omega_{\rm
o}+\omega_{\rm e}-\omega^{0}_{\rm p})$ and the state function
$\Phi$ for SPDC reduces to ${\tilde \chi}^{(2)}\left[\Delta({\bf
q},-{\bf q}; \omega, \omega_{\rm p}^0 - \omega)\right]$. Figure
1(a) shows the absolute square of the state function in
Eq.~(\ref{Chi-General}), the familiar ${\rm sinc}^2(L\Delta/2)$
distribution of SPDC from a single bulk crystal.

\subsubsection{Periodic Nonlinearity}

We now consider a medium of thickness $L$ with a periodic
distribution of nonlinear susceptibility
$\chi^{(2)}(z)=\chi^{(2)}(z+\Lambda)$ within the medium. Such
materials are widely used in classical nonlinear optics
\cite{fejer-PPLN} and have recently been employed for generation
of SPDC \cite{gisin-PPLN}. We may write

\begin{equation}\label{Chi-Periodic-General}
    \chi^{(2)}(z) = \chi_0
        g(z)\,{\rm rect}_{[-L,0]}(z)\,,
\end{equation}

\noindent where $g(z)$ can be expressed in the Fourier series

\begin{equation}\label{g-z}
  g(z)=\sum^{\infty}_{m=-\infty}G_{m}\ e^{{\rm i}K_{m}z}
\end{equation}

\noindent with $K_{m} \equiv 2\pi m/\Lambda$. The Fourier
Transform of Eq.~(\ref{Chi-Periodic-General}) is then given by

\begin{eqnarray}\label{FT-Chi-Periodic-General}
    {\tilde \chi}^{(2)}(\Delta)
        = \chi_0 L\sum^{\infty}_{m=-\infty}G_{m}\
        {\rm sinc}\left[\frac{L}{2}(\Delta+K_{m})\right]
                e^{{\rm -i}\frac{L}{2}(\Delta+K_{m})}\,.
\end{eqnarray}

For example, let us consider the case of a sinusoidal distribution
of nonlinear susceptibility with period $\Lambda$, for which

\begin{equation}\label{Chi-Periodic}
    \chi^{(2)}(z) = \chi_0
        \cos\left(\frac{2\pi}{\Lambda}z\right)\,{\rm rect}_{[-L,0]}(z)\,,
\end{equation}

\noindent which yields

\begin{eqnarray}\label{FT-Chi-Periodic-}
    {\tilde \chi}^{(2)}(\Delta)
        ~=~\chi_0L\left\{ {\rm
        sinc}\left[\frac{L}{2}\left(\Delta+\frac{2\pi}{\Lambda}\right)\right]
                e^{{\rm -i}\frac{L}{2}\left(\Delta+\frac{2\pi}{\Lambda}\right)}+
                     {\rm
                     sinc}\left[\frac{L}{2}\left(\Delta-\frac{2\pi}{\Lambda}\right)\right]
                e^{{\rm
                -i}\frac{L}{2}\left(\Delta-\frac{2\pi}{\Lambda}\right)}\right\}\,.
\end{eqnarray}

\noindent In this case, we obtain phase-matching conditions
similar to the first-order quasi-phase matching (QPM) observed in
periodically poled nonlinear crystals. The extra component $\pm
2\pi/\Lambda$ above is analogous to the grating vector in
first-order QPM. Figure 1(b) shows the absolute square of the
state function of the down-converted light obtained from a single
crystal with the nonlinearity profile given in
Eq.~(\ref{Chi-Periodic}).

\subsubsection{Cascaded Bulk Crystals Separated by Linear Media}

A simple example of a medium with an inhomogeneous distribution of
nonlinearity is a cascade of multiple bulk crystals separated by
linear dielectrics. Consider, for example, a cascade of $N$ bulk
crystals separated by $N-1$ linear media. Let each nonlinear
crystal $j$ have thickness $L_j$, constant nonlinearity
$\chi_{0_j}$, and separation distance $d_{j}$ from the previous
crystal. The overall nonlinear susceptibility of this system is
then given by

\begin{equation}\label{Chi-Cascade-N}
    \chi^{(2)}(z) = \sum^{N}_{j=1}\,\epsilon_{j}\chi_{0_j}\,{\rm
rect}_{[-L_{j},0]}\left[z+\sum_{k=j+1}^{N}(d_{k}+L_{k})\right]\,,
\end{equation}

\noindent where the terms of the summation inside the rect
function are taken to be zero if $k > N$. Here $\epsilon=\pm 1$
represents the sign of the quadratic susceptibility, which depends
on the orientation of the optical axis of the $j$-th crystal. Note
that in this equation, the $z=0$ point is placed at the output
plane of the last crystal. In such a configuration, the function
${\tilde \chi}^{(2)}(\Delta)$ in Eq.~(\ref{FT-Phi-General})
becomes

\begin{eqnarray}\label{FT-Chi-Cascade-N}
    {\tilde \chi}^{(2)}(\Delta)~=~
        \sum^{N}_{j=1}\,\epsilon_{j}\chi_{0_j}\,L_{j}\,{\rm
sinc}\left({{L_{j}\Delta_{j}}\over 2}\right)\,
            e^{{\rm -i}{{L_{j}\Delta_{j}}\over2}}\,
            e^{{\rm -i}\sum^{N}_{k=j+1}\,\left({{L_{k}\Delta_{k}}}+
                       {{d_{k}\Delta_{k}^{\prime}}}\right)}
\end{eqnarray}

\noindent where the wave vector mismatch function $\Delta_{j}$ is
independent of $z$ and $\Delta^{\prime}$. As seen from
Eq.~(\ref{Delta}), $\Delta^{\prime}$ depends on the dispersiveness
of the linear medium~\cite{bur97}.

We now consider the particular case of a cascade of two bulk
crystals of the same material [See Fig.~2(a)] separated by a
linear but dispersive medium. The explicit form of the
nonlinearity is given by

\begin{equation}\label{Chi-Cascade}
    \chi^{(2)}(z) = \chi_0\,{\rm rect}_{[-L_{1},0]}(z+d+L_{2})+
               \epsilon \,\chi_0\,{\rm rect}_{[-L_{2},0]}(z)\,,
\end{equation}

\noindent where $\epsilon=+1$ if the optical axes of the two
crystals are parallel and $\epsilon=-1$ if the optical axes are
antiparallel. For such a configuration

\begin{eqnarray}\label{FT-Chi-Cascade}
    {\tilde \chi}^{(2)}(\Delta)~=~
        {\chi_0} \,\left\{ L_{1}\,{\rm sinc}\left({{L_{1}\Delta}\over 2}\right)\,
            e^{{\rm -i}{{L_{1}\Delta}\over2}}\,
            e^{{\rm i}\left({{L_{2}\Delta}}+
                       {{d\Delta^{\prime}}}\right)}  +\,\epsilon\,
        L_{2}\,{\rm sinc}\left({{L_{2}\Delta}\over 2}\right)\,
        e^{{\rm -i}{{L_{2}\Delta}\over2}}\right\}\,.
\end{eqnarray}

\noindent The absolute square of the state function of SPDC in
this configuration is given in Fig.~2(a), where the envelope is
governed solely by the dispersion in the nonlinear crystals. The
period of the modulation inside this envelope is determined
primarily by the dispersion in the linear medium between the
crystals, while the amplitude of this modulation is determined by
the ratio of the crystal thicknesses. Figure 2(b) shows the
absolute square of the state function in the special case of two
bulk crystals of the same material with the same thickness. In
this condition the amplitude of the modulation inside the envelope
is maximized.

\subsection{Two-Photon Amplitude and Fourth-Order Correlation}

We now consider the propagation of the down-converted light
through an arbitrary linear optical system to a pair of detectors,
as illustrated in Fig.~3. The joint probability amplitude of
detecting the photon pair at the space-time coordinates $({\mathbf
x}_{\rm A},t_{\rm A})$ and $({\mathbf x}_{\rm B},t_{\rm B})$ is
given by

\begin{equation}\label{Biphoton-Definition}
    A({\mathbf x}_{\rm A},t_{\rm A};{\mathbf x}_{\rm B},t_{\rm B})=
            \langle 0| \hat{E}^{(+)}_{\rm A}({\mathbf x}_{\rm A},t_{\rm A})
                       \hat{E}^{(+)}_{\rm B}({\mathbf x}_{\rm B},t_{\rm B})
 |\Psi^{(2)} \rangle
\end{equation}

\noindent where $E_{\rm A}^{(+)}$ and $E_{\rm B}^{(+)}$ are the
positive-frequency components of the electric fields at points A
and B. The explicit forms of the quantum fields present at the
detection locations are given by

\begin{eqnarray}\label{Field-Detector-General}
    \hat E_{\rm A}^{(+)}({\bf x}_{\rm A},t_{\rm A})=
        \sum_{j={\rm e},{\rm o}}\int  d{\bf q}\,d\omega~e^{{\rm -i}\omega t_{\rm A}}\
        {\cal H}_{\rm A \it j}({\bf x}_{\rm A},{\bf q};\omega)\
               \hat a_{j}({\bf q},\omega)\,,\nonumber\\
    \hat E_{\rm B}^{(+)}({\bf x}_{\rm B},t_{\rm B})=
        \sum_{j={\rm e},{\rm o}}\int  d{\bf q}\,d\omega~e^{{\rm -i}\omega t_{\rm B}}\
        {\cal H}_{\rm B \it j}({\bf x}_{\rm B},{\bf q};\omega)\
                \hat a_{j}({\bf q},\omega)\,,
\end{eqnarray}

\noindent where the transfer function ${\cal H}_{ij}$ ($i={\rm
A},{\rm B}$ and $j={\rm e},{\rm o})$ describes the propagation of
a mode $({\bf q},\omega)$ through the optical system from the
output plane of the nonlinear medium to the detection plane.
Substitution of Eqs.~(\ref{Psi-General}) and
(\ref{Field-Detector-General}) into
Eq.~(\ref{Biphoton-Definition}) yields a general form for the
two-photon detection probability amplitude,

\begin{eqnarray}\label{Biphoton-General}
    A({\mathbf x}_{\rm A},t_{\rm A};{\mathbf x}_{\rm B},t_{\rm B})=
        A_{{\rm Ao},{\rm Be}}({\mathbf x}_{\rm A},t_{\rm A};{\mathbf x}_{\rm B},t_{\rm B})+
        A_{{\rm Bo},{\rm Ae}}({\mathbf x}_{\rm A},t_{\rm A};{\mathbf x}_{\rm B},t_{\rm B})
\end{eqnarray}

\noindent where the probability amplitude $A_{{\rm Ao},{\rm Be}}$
for finding the signal photon in arm A and the idler photon in arm
B is defined as
\begin{eqnarray}\label{A2-i-oe}
    A_{{\rm Ao},{\rm Be}}({\mathbf x}_{\rm A},t_{\rm A};{\mathbf x}_{\rm B},t_{\rm B})=
        \int d{\bf q}_{\rm o}d{\bf q}_{\rm e}\,d\omega_{\rm o}d\omega_{\rm e}\
        \Phi({\bf q}_{\rm o},{\bf q}_{\rm e};\omega_{\rm o},\omega_{\rm e})\
                e^{-{\rm i}(\omega_{\rm o}t_{\rm A}+\omega_{\rm e}t_{\rm B})}
                        \nonumber\\\times
        {\cal H}_{\rm Ao}({\bf x}_{\rm A},{\bf q}_{\rm o};\omega_{\rm o})\
        {\cal H}_{\rm Be}({\bf x}_{\rm A},{\bf q}_{\rm e};\omega_{\rm e})
\end{eqnarray}
and $A_{{\rm Bo},{\rm Ae}}({\mathbf x}_{\rm A},t_{\rm A};{\mathbf
x}_{\rm B},t_{\rm B})$ is obtained by exchanging the indices ${\rm
A}\leftrightarrow {\rm B}$.

The joint probability density for detection of the signal and
idler photons at space-time points $({\mathbf x}_{\rm A},t_{\rm
A})$ and $({\mathbf x}_{\rm B},t_{\rm B})$ is given by the
fourth-order correlation function, is given by the absolute square
of Eq.~(\ref{Biphoton-General}):
\begin{eqnarray}\label{Fourth-Order-Correlation}
  G^{(2)}({\mathbf x}_{\rm A},t_{\rm A};{\mathbf x}_{\rm B},t_{\rm B})
        &=&|A_{{\rm Ao},{\rm Be}}({\mathbf x}_{\rm A},t_{\rm A};{\mathbf x}_{\rm B},t_{\rm B})|^{2}+
         |A_{{\rm Bo},{\rm Ae}}({\mathbf x}_{\rm A},t_{\rm A};{\mathbf x}_{\rm B},t_{\rm B})|^{2}
                \nonumber\\&&+
            2\, \Re e[
         A_{{\rm Ao},{\rm Be}}^{\ast}({\mathbf x}_{\rm A},t_{\rm A};{\mathbf x}_{\rm B},t_{\rm B})\
         A_{{\rm Bo},{\rm Ae}}({\mathbf x}_{\rm A},t_{\rm A};{\mathbf x}_{\rm B},t_{\rm B})].
\end{eqnarray}
With current technology, quantum interferometry is performed using
slow detectors that cannot resolve signals on the characteristic
time scale of down-conversion (the inverse of down-conversion
bandwidth), which is typically less than 1 ps. In addition, the
detectors used in our experiments have a large active area
compared to the width of the SPDC beams at the detection planes.
Under these conditions, the coincidence count rate $R$ is readily
expressed in terms of the two-photon detection probability
amplitude $A$ by integrating the fourth-order correlation function
$G^{(2)}({\mathbf x}_{\rm A},t_{\rm A};{\mathbf x}_{\rm B},t_{\rm
B})$ over all space and time,

\begin{eqnarray}\label{Rate-int}
  R=\int dt_{\rm A} dt_{\rm B} \, d{\mathbf x}_{\rm A}
            d{\mathbf x}_{\rm B}\
        |A({\mathbf x}_{\rm A},t_{\rm A};{\mathbf
                  x}_{\rm B},t_{\rm B})|^{2}\,.
\end{eqnarray}

This expression for the count rate can be separated into two terms
as

\begin{equation}\label{Rate-as-sum}
  R=R_{0}+R_{\rm int}\,,
\end{equation}

\noindent where the baseline term is

\begin{eqnarray}\label{R-AoBe}
  R_{0}=&
    \int d\omega^{\prime}d\omega\,
     d{\bf q}_{\rm o} d{\bf q}_{\rm e}\, d{\bf q}_{\rm o}^{\prime}d{\bf q}_{\rm e}^{\prime}\
       & \Phi({\bf q}_{\rm o},{\bf q}_{\rm e};\omega,\omega^{\prime})\
        \Phi^{\ast}({\bf q}_{\rm o}^{\prime},{\bf q}_{\rm e}^{\prime};\omega,\omega^{\prime})\ \nonumber \\
   && \times\, \left[{\mathcal S}_{\rm AB}({\bf q}_{\rm o},{\bf q}_{\rm e},{\bf q}_{\rm o}^{\prime},{\bf
    q}_{\rm e}^{\prime};\omega,\omega^{\prime})+{\mathcal S}_{\rm BA}({\bf q}_{\rm o},{\bf q}_{\rm e},{\bf q}_{\rm o}^{\prime},{\bf
    q}_{\rm e}^{\prime};\omega,\omega^{\prime})\right]
\end{eqnarray}

\noindent and the interference term is

\begin{eqnarray}\label{Interf}
  R_{\rm int}=2 \Re e
    \int d\omega^{\prime}d\omega\,
     d{\bf q}_{\rm o}\ d{\bf q}_{\rm e}\, d{\bf q}_{\rm o}^{\prime}d{\bf q}_{\rm e}^{\prime}\
        \Phi({\bf q}_{\rm o},{\bf q}_{\rm e};\omega^{\prime},\omega)\
        \Phi^{\ast}({\bf q}_{\rm o}^{\prime},{\bf q}_{\rm e}^{\prime};\omega,\omega^{\prime})\
    {\mathcal S}_{\rm AB}({\bf q}_{\rm e},{\bf q}_{\rm o},{\bf q}_{\rm o}^{\prime},{\bf
    q}_{\rm e}^{\prime};\omega,\omega^{\prime})\,.
\end{eqnarray}

\noindent In Eqs.~(\ref{R-AoBe}) and (\ref{Interf}) the state
function $\Phi$ weights the signal and idler modes in the process
of generation, while the function ${\mathcal S}_{AB}$ weights
these modes in the process of propagation through the optical
system. Explicitly,

\begin{equation}\label{Spacial-integral}
  {\mathcal S}_{\rm AB}({\bf q}_{\rm e},{\bf q}_{\rm o},{\bf q}_{\rm o}^{\prime},{\bf q}_{\rm e}^{\prime};\omega,\omega^{\prime})=
            \left\langle {\cal H}^{\ast}_{\rm Ao}({\bf x}_{\rm A},{\bf q}_{\rm o}^{\prime};\omega)\
                    {\cal H}_{\rm Ao}({\bf x}_{\rm A},{\bf q}_{\rm o};\omega)\right\rangle_{{\bf x}_{\rm A}}
            \left\langle {\cal H}^{\ast}_{\rm Be}({\bf x}_{\rm B},{\bf q}_{\rm e}^{\prime};\omega^{\prime})\
                    {\cal H}_{\rm Be}({\bf x}_{\rm B},{\bf q}_{\rm e};\omega^{\prime})\right\rangle_{{\bf x}_{\rm B}}
\end{equation}

\noindent where $\langle\cdot\rangle_{{\bf x}_{i}}$ indicates
integration over the total detector area.

Note from Eqs.~(\ref{Biphoton-General}) and (\ref{A2-i-oe}) that
the two-photon detection probability amplitude is completely
specified by ${\cal H}_{ij}$ ($i={\rm A},{\rm B}$ and $j={\rm
o},{\rm e}$), $\Phi({\bf q}_{\rm o},{\bf q}_{\rm e};\omega_{\rm
o},\omega_{\rm e})$, and the physical location of detectors A and
B. As we have seen in Eq.~(\ref{FT-Phi-General}), we may control
the structure of the state function $\Phi$ by a judicious choice
of the pump spectral profile, the longitudinal distribution of
nonlinearity, and the dispersion in the crystal. We may further
control the two-photon detection amplitude, and hence the
quantum-interference pattern, through the choice of the optical
system. Note that states with different state functions can lead
to the same quantum-interference pattern through an appropriate
design of the optical system.

In the experimentally relevant case of a monochromatic plane wave
pump field, Eqs.~(\ref{R-AoBe}) and~(\ref{Interf}) become

\begin{eqnarray}\label{R-AoBe-CW}
  R_{0}&=
    \int d\omega\,
    \int d{\bf q}\ d{\bf q}^{\prime}\
       & {\tilde \chi}^{(2)\ast}[\Delta({\bf q}^{\prime},-{\bf q}^{\prime};\omega,\omega^{0}_{\rm p}-\omega)]\
        {\tilde \chi}^{(2)}[\Delta({\bf q},-{\bf
        q};\omega,\omega^{0}_{\rm p}-\omega)] \nonumber \\
    && \times\,\left[{\bar {\mathcal S}}_{\rm AB}({\bf q},{\bf
q}^{\prime};\omega)+{\bar {\mathcal S}}_{\rm BA}({\bf q},{\bf
    q}^{\prime};\omega)\right]
\end{eqnarray}

\noindent and

\begin{eqnarray}\label{Interf-CW}
  R_{\rm int}=2 \Re e
    \int d{\bf q}\ d{\bf q}^{\prime}\
        {\tilde \chi}^{(2)\ast}[\Delta({\bf q}^{\prime},-{\bf q}^{\prime};\omega^{0}_{\rm p}-\omega,\omega)]\
        {\tilde \chi}^{(2)}[\Delta({\bf q},-{\bf q};\omega,\omega^{0}_{\rm p}-\omega)]\
    {\bar {\mathcal S}}_{\rm AB}({\bf q},-{\bf q}^{\prime};\omega)
\end{eqnarray}

\noindent where we use the shorthand

\begin{equation}\label{Spacial-integral-CW}
    {\bar {\mathcal S}}_{\rm AB}({\bf q},{\bf q}^{\prime};\omega)=
    {\mathcal S}_{\rm AB}({\bf q},-{\bf q},{\bf q}^{\prime},-{\bf
    q}^{\prime};\omega,\omega^{0}_{\rm p}-\omega).
\end{equation}

\noindent Thus we see that for a monochromatic plane wave pump,
the quantum-interference pattern is critically dependent on the
form of ${\tilde \chi}^{(2)}(\Delta)$, which we are free to choose
as a design parameter~\cite{fejer-PPLN}. In principle, the only
limitation on the class of amplitudes $A$ which we are able to
prepare with this method is the restriction that the optical
system is linear.

\subsection{Quantum Interference with a Cascaded Pair of Bulk Crystals}

We now apply the above formalism to the case of two cascaded bulk
crystals separated by a dispersive but linear dielectric medium
such as an air gap. For simplicity, we again consider the medium
to be pumped by a monochromatic plane wave. Owing to the structure
of the nonlinearity for this particular case
[Eq.~(\ref{Chi-Cascade})], the overall two-photon detection
probability amplitude is the sum of the two amplitudes associated
with each single crystal~\cite{bur97,ata01}. Each of the
amplitudes in Eq.~(\ref{Biphoton-General}) can then be written as

\begin{eqnarray}\label{A2-ampl}
   A_{{\rm Ao},{\rm Be}}({\mathbf x}_{\rm A},t_{\rm A};{\mathbf x}_{\rm B},t_{\rm B})=
            A_{{\rm Ao},{\rm Be}}^{(1)}({\mathbf x}_{\rm A},t_{\rm A};{\mathbf x}_{\rm B},t_{\rm B})+\epsilon\
            A_{{\rm Ao},{\rm Be}}^{(2)}({\mathbf x}_{\rm A},t_{\rm A};{\mathbf x}_{\rm B},t_{\rm B})
\end{eqnarray}

\noindent where $\epsilon=\pm 1$ as in Eq.~(\ref{Chi-Cascade}),
and a similar expression for $A_{\rm Bo,Ae}$ is obtained by
exchanging the indices A $\leftrightarrow$ B. $A^{(r)}_{{\rm
Ao},{\rm Be}}$ for $r=(1,2)$ is the probability amplitude of
finding the o-polarized photon generated in the $r$-th crystal in
arm A and the e-polarized photon from the $r$-th crystal in arm B.
From Eq.~(\ref{A2-i-oe}),

\begin{eqnarray}\label{A2-i-oe-r}
    A^{(r)}_{{\rm Ao},{\rm Be}}({\mathbf x}_{\rm A},{\mathbf x}_{\rm B};t)&=&
        \int d\nu\, d{\mathbf q}\,
            e^{-i\nu t} {\tilde \chi}_r^{(2)}({\mathbf q},\nu)\
            {\mathcal H}_{\rm Ao}({\mathbf x}_{\rm A};{\mathbf q},\nu)\
            {\mathcal H}_{\rm Be}({\mathbf x}_{\rm B};-{\mathbf q},-\nu)
\end{eqnarray}

\noindent where the angular frequency $\nu = \omega - \omega_{\rm
p}^0 / 2$ is the deviation from the central frequency $\omega_{\rm
p}^0 / 2$, $t=t_{\rm A}-t_{\rm B}$ is the time difference between
detection events, and ${\tilde \chi}_{r}^{(2)}({\mathbf q},\nu)$
is the inverse Fourier transform of the nonlinearity profile of
the $r$-th crystal. $A^{(r)}_{{\rm Bo},{\rm Ae}}$ is likewise
obtained by a suitable exchange of the indices. Note that we have
omitted an overall phase factor $\exp[-i\omega^{0}_{\rm p} (t_{\rm
A}+t_{\rm B})/2]$ which appears outside the integral in
Eq.~(\ref{A2-i-oe-r}), since in experimental practice we are
interested only in the absolute square $|A({\mathbf x}_{\rm
A},t_{\rm A};{\mathbf x}_{\rm B},t_{\rm B})|^{2}$, so this factor
does not introduce any relative phase between the terms of
Eq.~(\ref{Biphoton-General}).

As we are considering bulk crystals, the nonlinearity profile of
each crystal $r$ is uniform, and thus
\begin{eqnarray}\label{zeta-r-simple}
  {\tilde \chi}^{(2)}_{r}({\mathbf q},\nu)=
           \chi_0 Q_{r}\left({\mathbf q},\nu\right)\cdot
     \int dz\
            {\rm rect}_{[-L_{r},0]}(z)\
            e^{i\Delta_{r}({\mathbf q},\nu)z}\,,
\end{eqnarray} where

\begin{equation}\label{P-d-i}
 Q_{r}\left({\mathbf q},\nu\right)=
     e^{-i[d\Delta^{\prime}({\mathbf q},\nu)+L_{2}\Delta_{2}
        ({\mathbf q},\nu)]\ \delta_{r,1}}
\end{equation}

\noindent is the transfer function for propagation of the signal
(o-polarized) and idler (e-polarized) fields generated in the
first crystal through the linear dispersive medium of thickness
$d$ and thence through the second crystal of thickness $L_{2}$.
Alternatively, $Q_r \left({\mathbf q},\nu\right)$ may be thought
of as the phase accumulated in the shift of the rect function for
the first crystal by a distance $-(L_2 + d)$. As given in
Eq.~(\ref{Delta}), $\Delta_{r}$ is the wave vector mismatch
function due to dispersion in the $r$-th crystal and
$\Delta^{\prime}$ is the wave vector mismatch function due to
dispersion in the linear medium. The symbol $\delta_{r,1}$
represents the Kronecker delta where $\delta_{1,1}=1$ and
$\delta_{2,1}=0$.

\subsubsection{Coincidence Detection}

Taking the absolute square of Eq.~(\ref{A2-ampl}) gives
interference between the probability amplitudes of finding a pair
generated in the first crystal and finding a pair generated in the
second crystal. Indeed, substitution of
Eqs.~(\ref{Biphoton-General}) and~(\ref{A2-ampl}) into
Eq.~(\ref{Rate-int}) gives the coincidence count rate as a sum of
three contributions

\begin{equation}\label{Rate-simpl-2crys}
  R=R^{(1)}+R^{(2)}+R^{(12)}
\end{equation}

\noindent where the first two terms are the coincidence-count
rates for single-crystal SPDC, and the last term arises as
interference between the two single-crystal amplitudes. Recalling
Eq.~(\ref{Rate-as-sum}), each term in Eq.~(\ref{Rate-simpl-2crys})
can in turn be broken down into baseline and interference terms

\begin{equation}\label{Rate-simpl-one-crystal}
  R^{(h)}=R_{0}^{(h)}+R_{\rm int}^{(h)}
\end{equation}

\noindent where $h=1,2,12$.

We now use this theory to predict quantum-interference patterns in
the interferometer shown in Fig.~4. In this system, the transfer
function ${\mathcal H}_{ij}$ is separable into
diffraction-dependent and -independent factors as

\begin{equation}\label{H-T-H}
  {\mathcal H}_{ij}({\mathbf x}_{j};{\mathbf q},\nu)=H_{i}({\mathbf x}_{i};{\mathbf q},\nu)\,{\mathcal T}_{ij}\,e^{i\kappa_{j}({\mathbf q},\nu) l_{\tau}}
\end{equation}

\noindent where the polarization-independent components are
grouped into $H_i$ and the remainder are grouped into ${\mathcal
T}_{ij}\, e^{i\kappa_{j}({\mathbf q},\nu) l_{\tau}}$. In this
case, ${\mathcal T}_{ij}=({\mathbf e}_{i}\cdot{\mathbf e}_{j})$ is
the projection of the unit photon polarization vector ${\mathbf
e}_{j}$ ($j={\rm o},{\rm e}$) onto the axis of the polarization
analyzer in front of detector $i=({\rm A},{\rm B})$, and the
exponential factor is the transfer function of the delay line.

The delay line, which is often treated as a simple phase shift, is
a dispersive optical element which may alter the spatial and/or
the spectral profile of the two-photon probability amplitude.
Experimentally, the delay line consists of a birefringent quartz
plate of variable thickness, modelled by the propagation function
$e^{i\kappa_{j}({\mathbf q},\nu) l_{\tau}}$, where the
longitudinal projections $\kappa_{j}$ of the signal and idler wave
vectors are defined in Eq.~(\ref{kz-int}) and $l_\tau$ is the
thickness of the birefringent plate, which induces a relative
optical-path delay $\tau$. This propagation function, while here
used to describe the delay line, is technically a valid transfer
function for any nonabsorbing dispersive optical element.

Given this particular optical system, then, the single-crystal
coincidence-count rates are given by

\begin{eqnarray}\label{R-Inf-r-2crys}
  R^{(r)}_{0} =&& \int d\nu\, d{\mathbf q} d{\mathbf
                             q}^{\prime}\,
        {\mathcal F}_{\rm AB}({\mathbf q},{\mathbf q}^{\prime},\nu)\
        e^{-{\rm i}[\eta_{\tau}({\mathbf q},\nu)-\eta_{\tau}({\mathbf
q}^{\prime},\nu)]}
                    \nonumber\\&&\times \,
      \left[\mu_{{\rm Ao},{\rm Be}}^{2}\
        {\tilde \chi}^{(2) \ast}_r({\mathbf q}^{\prime},\nu)\
        {\tilde \chi}^{(2)}_r({\mathbf q},\nu)+\mu_{{\rm Bo},{\rm Ae}}^{2}\
        {\tilde \chi}^{(2) \ast}_r(-{\mathbf q}^{\prime},-\nu)\
        {\tilde \chi}^{(2)}_r(-{\mathbf q},-\nu)
     \right]
\end{eqnarray}

\noindent and

\begin{eqnarray}\label{R-0-r-2crys}
    R^{(r)}_{\rm int} = &&
        \int d\nu\, d{\mathbf q} d{\mathbf q}^{\prime}\,
        {\mathcal F}_{\rm AB}({\mathbf q},-{\mathbf q}^{\prime},\nu)\
        e^{-{\rm i}[\eta_{\tau}({\mathbf q},\nu)-\eta_{\tau}({\mathbf
q}^{\prime},-\nu)]}
                    \nonumber\\&&\times \,
        \mu_{{\rm Ao},{\rm Be}}\ \mu_{{\rm Bo},{\rm Ae}}
     \left[
        {\tilde \chi}^{(2) \ast}_r({\mathbf q}^{\prime},-\nu)\
        {\tilde \chi}^{(2)}_r({\mathbf q},\nu)+
        {\tilde \chi}^{(2) \ast}_r(-{\mathbf q}^{\prime},\nu)\
        {\tilde \chi}^{(2)}_r(-{\mathbf q},-\nu)
     \right]
\end{eqnarray}

\noindent where $r=1,2$ is the crystal index,
$\mu_{ij,lm}={\mathcal T}_{ij}\cdot {\mathcal T}_{lm}$ ($i,l={\rm
A},{\rm B}$ and $j,m={\rm e},{\rm o}$) is the projection of the
polarization of the $i$-th photon onto the $j$-th basis
polarization and the polarization of the $l$-th photon onto the
$m$-th basis polarization, and the phase function

\begin{equation}\label{delay-line}
  \eta_{\tau}({\mathbf q},\nu)=-
            \left[\kappa_{\rm o}\left(\nu,{\mathbf q}\right)+
            \kappa_{\rm e}\left(-\nu, -{\mathbf
            q}\right)\right]l_{\tau}
\end{equation}

\noindent depends on the dispersion introduced by the delay line.
The integral over the detection planes

\begin{eqnarray}\label{A-F-D1}
 {\mathcal F}_{\rm AB}({\mathbf q},\pm {\mathbf q}^{\prime},\nu)&=&
  \left\langle
      H_{\rm A}\left({\mathbf x}_{\rm A},{\mathbf q};\nu\right)\
      H^{\ast}_{\rm A}\left({\mathbf x}_{\rm A},\pm{\mathbf q}^{\prime};\nu\right)\
  \right\rangle_{{\mathbf x}_{\rm A}}
  \left\langle
      H_{\rm B}\left({\mathbf x}_{\rm B},-{\mathbf q};-\nu\right)\
      H^{\ast}_{\rm B}\left({\mathbf x}_{\rm B},\mp{\mathbf q}^{\prime};-\nu\right)\
  \right\rangle_{{\mathbf x}_{\rm B}}
\end{eqnarray}

\noindent is an analog of the function ${\bar {\mathcal S}}_{\rm
AB}$ for the polarization-independent elements of the system only.

Meanwhile, the coincidence-count rates which arise collectively
between the contributions from the two crystals are given by

\begin{eqnarray}\label{R-Inf-12-2crys}
  R^{(12)}_{0}=&& \epsilon\ \int d\nu\, d{\mathbf q} d{\mathbf
  q}^{\prime}\,
        {\mathcal F}_{\rm AB}({\mathbf q},{\mathbf q}^{\prime},\nu)\
        e^{-{\rm i}[\eta_{\tau}({\mathbf q},\nu)-\eta_{\tau}({\mathbf
q}^{\prime},\nu)]}
                        \nonumber \\&&\times \,
     \sum_{r=1,2}\left[\mu_{{\rm Ao},{\rm Be}}^{2}\
        {\tilde \chi}^{(2) \ast}_r({\mathbf q}^{\prime},\nu)\
        {\tilde \chi}^{(2)}_{3-r}({\mathbf q},\nu)+\mu_{{\rm Bo},{\rm Ae}}^{2}\
        {\tilde \chi}^{(2) \ast}_r(-{\mathbf q}^{\prime},-\nu)\
        {\tilde \chi}^{(2)}_{3-r}(-{\mathbf q},-\nu)\right]
\end{eqnarray}

\noindent and

\begin{eqnarray}\label{R-0-12-2crys}
  R^{(12)}_{\rm int} = &&\epsilon\ \int d\nu\, d{\mathbf q}
d{\mathbf
  q}^{\prime}\,
        {\mathcal F}_{\rm AB}({\mathbf q},-{\mathbf q}^{\prime},\nu)\
        e^{-{\rm i}[\eta_{\tau}({\mathbf q},\nu)-\eta_{\tau}({\mathbf
q}^{\prime},-\nu)]}
                        \nonumber \\&&\times \,
        \mu_{{\rm Ao},{\rm Be}}\ \mu_{{\rm Bo},{\rm Ae}}
    \sum_{r=1,2}\left[
        {\tilde \chi}^{(2) \ast}_r({\mathbf q}^{\prime},-\nu)\
        {\tilde \chi}^{(2)}_{3-r}({\mathbf q},\nu)+
        {\tilde \chi}^{(2) \ast}_r(-{\mathbf q}^{\prime},\nu)\
        {\tilde \chi}^{(2)}_{3-r}(-{\mathbf q},-\nu)\right]\,.
\end{eqnarray}

By combining Eqs.~(\ref{R-Inf-r-2crys}), (\ref{R-0-r-2crys}),
(\ref{R-Inf-12-2crys}), and (\ref{R-0-12-2crys}), the overall
coincidence-count rate $R(\tau)$ can be organized into a general
form

\begin{equation}\label{R-final-2crys}
  R(\tau)= R_{0}\left[1+v_{\rm pol}V(\tau)\right]
\end{equation}

\noindent where $R_{0}$ is the baseline coincidence-count rate and
the overall projection of both photon polarizations onto the basis
of the polarization analyzers is given by the factor

\begin{equation}\label{v-pol}
  v_{\rm pol}=2\frac{\mu_{{\rm Ao},{\rm Be}}\ \mu_{{\rm Bo},{\rm Ae}}}{\mu_{{\rm Ao},{\rm Be}}^{2}+\mu_{{\rm Bo},{\rm Ae}}^{2}}\,.
\end{equation}

\noindent Observe that the $\tau$-dependence of the quantum
interference pattern is then contained solely in the visibility
function $V(\tau)$.

\subsubsection{Cascade of Two Identical Crystals}

In the experiments presented in this paper, the two crystals are
of the same material and have equal thicknesses $L$. The apertures
are symmetric for both transverse directions, and the analyzers
are set $45^\circ$ from the optical axis, so $v_{\rm pol} = -1$.
No spectral filters are used.  For these conditions, the explicit
form of the visibility function in Eq.~(\ref{R-final-2crys})
becomes


\begin{eqnarray}\label{V2-general}
  V(\tau)&=&\frac{1}{1+\rho^2} \int dz\
                \Pi_{L}(z)\
                \Pi_{L}\left(\frac{2\tau}{D}-2L-z\right)\
                     {\mathcal G}^{(1)}\left(\frac{z}{L};\frac{\tau}{LD}\right)\nonumber\\
              &&  +\frac{1}{1+\rho^2}\int dz\
                \Pi_{L}(z)\
                \Pi_{L}\left(\frac{2\tau}{D}-z\right)\
             {\mathcal
             G}^{(2)}\left(\frac{z}{L};\frac{\tau}{LD}\right)\nonumber\\
              &&  +2\epsilon \frac{\rho}{1+\rho^2}\int dz\
                \Pi_{L}(z)\
                \Pi_{L}\left(\frac{2\tau}{D}-L-z\right)\
                     \Re e\left[{\mathcal G}^{(12)}\left(\frac{z}{L}+\frac{1}{2};\frac{\tau}{LD}\right)
                                  \ e^{-i\Delta^{\prime} d}\right]
\end{eqnarray}

\noindent where ${\mathcal G}^{(h)}$ is defined in
Appendix~\ref{App-2}, $\rho=(d_{1}+d)/d_{1}$, $\Pi_{L}(z)$ is the
unit rect function from $[0,L]$, and $D = u_{\rm o}^{-1} - u_{\rm
e}^{-1}$ is the dispersion coefficient of the nonlinear medium. It
is through the ${\mathcal G}$-functions that spatiotemporal
effects enter the quantum interference pattern. Details on the
derivation of Eq.~(\ref{V2-general}) can be found in
Appendix~\ref{App-2}.

The collective interference term in Eq.~(\ref{V2-general}) shows
interesting behavior in certain limits of crystal separation. If
the optical axes of the two crystals are parallel, the
coincidence-count rate reduces to that from a single crystal of
thickness $2L$ as $d\rightarrow 0$. Further, it reduces to that
from a single crystal of thickness $L$ as $d\rightarrow \infty$.
We also note the absence of any shoulder modulation with $\tau$,
an important indication of the purity of the polarization
Bell-state which is formed in postselection. In the case where the
two crystals have equal thickness, the strongest interference
occurs at delay $\tau=LD$. The visibility at this point is given
by
\begin{eqnarray}\label{V2-maximum}
  V(LD)=2\epsilon\frac{\rho}{1+\rho^{2}}
                     \int^{1}_{0} d\zeta\
                     \Re{\it e}\left[{\mathcal
                     G}^{(12)}\left(\zeta;1\right)\right]
\end{eqnarray}

\noindent where $\zeta=z/L$ is a convenient dimensionless
variable.

\subsubsection{Small-Aperture Approximation}

In the limit of very small apertures, no transverse wave vectors
are allowed to propagate through the interferometer, and in the
case of sufficiently thin bulk crystals and small separation
distances, the quantum interference is effectively described by
the conventionally used single-mode
theory~\cite{ata01,Sasha-theory}.

Figure 5 is a sketch illustrating how quantum interference arises
in the interferometer of Fig.~4 assuming two identical crystals of
thickness $L_{1}=L_{2}=L$, dispersion coefficient $D$, parallel
optical axes, and both polarization analyzers set to $45^\circ$.
In the limit of sufficiently small apertures, we may apply the
conventional single-mode theory \cite{Sasha-theory} and write the
third term in Eq.~(\ref{Fourth-Order-Correlation}) as the product
of two probability amplitudes $A(t_{\rm A} - t_{\rm B})$ and
$A^*(- t_{\rm A} + t_{\rm B})$, which slide back and forth across
the $t_{\rm A} - t_{\rm B}$ axis as the relative optical-path
delay $\tau$ is varied. In the diagonal portion of the
illustration, these amplitudes are depicted by two grey-and-white
rectangles. When the delay is set such that the two rectangles
overlap, interference can be seen.

Within each rectangle, the white box represents the probability
amplitude for detecting a photon pair produced in the first
crystal, while the grey box represents the detection amplitude for
a pair produced in the second. As the delay is set in the region
$0 \leq \tau \leq LD/2$, the grey boxes overlap (shown as black)
but the white boxes do not. In this regime, interference typical
of a single crystal of thickness $2L$ is observed. As the delay is
increased into the region $LD/2 \leq \tau \leq 3LD/2$, the photon
pairs produced in the first crystal become indistinguishable with
the photon pairs produced in the second crystal at the detection
planes. As such, the probability amplitudes of detecting photon
pairs produced in each each crystal exhibit collective
interference. This is seen pictorially by the overlap of the grey
boxes with the white boxes. Note that the two rectangles overlap
completely at $\tau = LD$, the center of the region of
interference. The interference visibility in this region depends
on the phase shift between these two amplitudes, which is in turn
dependent on spatial effects and the dispersion in the linear
medium separating the crystals. As the distance $d$ between
crystals is changed, the visibility in this region modulates
sinusoidally between $\pm 1$. As the delay is increased still
further into the region $3LD/2 \leq \tau \leq 2LD$, the white
boxes overlap (shown as light grey) but the grey ones do not, and
we again return to the regime of single-crystal interference. In
this way we can trace out the interference dip predicted by
Eq.~(\ref{V2-general}). This is illustrated in the inset at the
lower right while representative experimental data is shown in the
upper left inset. Figure~6 is similar to Fig.~5, but for the case
of unequal crystal thicknesses $L_1 \neq L_2$.

In this small-aperture approximation, the dispersion-related
${\mathcal G}$-functions in Eq.~(\ref{V2-general}) for the cases
of parallel (p) and antiparallel (a) optical axes at $\tau/LD=1$
are found in Appendix \ref{App-2} to be
\begin{equation}\label{G-sm-parallel}
  {\mathcal G}_{\rm p}^{(12)}(\zeta;1)=\exp (-i\Delta^{\prime} d)\
                                    \exp \left\{-i\frac{k_{\rm p}|{\mathbf M} L|^{2}}{8}\left[\frac{(\zeta-1)^{2}}{d_{1}}-
                                            \frac{(\zeta+1)^{2}}{d_{1}+d}\right]\right\}
\end{equation}

\noindent and

\begin{equation}\label{G-sm-antiparallel}
  {\mathcal G}_{\rm a}^{(12)}(\zeta;1)=\exp (-i\Delta^{\prime} d)\
                                    \exp \left\{-i\frac{k_{\rm p}|{\mathbf M}
L|^{2}}{8}\left[\frac{(\zeta-1)^{2}}{d_{1}}-
\frac{(\zeta-1)^{2}}{d_{1}+d}\right]\right\}\,,
\end{equation}

\noindent where $k_{\rm p}=2\pi/\lambda_{\rm p}$ and $|{\mathbf
M}| =\frac{\partial}{\partial \theta_{\rm e}} \ln[n_{\rm
e}(\omega^{0}_{\rm p}/2,\theta_{\rm p})]$ is the spatial walk-off
vector~\cite{rubin}.

When a plane wave pump is normally incident on thin crystals and
the apertures are sufficiently small, the strictly collinear
signal and idler beams are selected by the optical system, and no
spatial effects due to transverse wave vectors can be observed. In
this limit the dominant contribution to the phase between
probability amplitudes arises from dispersion in the linear medium
between the crystals [see Eqs.~(\ref{G-sm-parallel})
and~(\ref{G-sm-antiparallel})]. For degenerate SPDC this phase
term in air is~\cite{CRC}

\begin{equation}\label{air-dispersion}
  \phi_{\rm disp}(d)\equiv\Delta^{\prime}d=
            k_{\rm p}[n(\lambda_{\rm p})-n(2\lambda_{\rm p})]\ d \sim
            \pi\ 0.059\ d[{\rm mm}]
\end{equation}
where we have taken $\lambda_{\rm p}=351.1$~nm in keeping with our
experiments.

However, when long crystals are used, or the separation distance
between the two crystals becomes large, some spatial effects due
to transverse wave vectors become observable. The visibility at
the center of the region of interference $tau = LD$
[Eq.~(\ref{V2-maximum})] for the case of parallel optical axes is
given by
\begin{equation}\label{V2-maximum-general-sm-parallel}
  V_{\rm p}(LD)=2\frac{d_{1}(d_{1}+d)}{d_{1}^{2}+(d_{1}+d)^{2}}
                     \int^{1}_{0} d\zeta\
                     \cos\left\{\frac{k_{\rm p}|{\mathbf M}L|^{2}}{2(d_{1}+d)}
                     \left[(1+\zeta^{2})\frac{d}{4d_{1}}-\zeta\left(1+\frac{d}{2d_{1}} \right)\right]+\phi_{\rm
                     disp}(d)\right\}\,,
\end{equation}
while for the case of antiparallel optical axes, it is given by
\begin{equation}\label{V2-maximum-general-sm-antiparallel}
  V_{\rm a}(LD)=-2\frac{d_{1}(d_{1}+d)}{d_{1}^{2}+(d_{1}+d)^{2}}
                     \int^{1}_{0} d\zeta\
                     \cos\left[\frac{k_{\rm p}|{\mathbf M}L|^{2}}{2(d_{1}+d)}\frac{d}{d_{1}}
                     \left(\zeta-1\right)^{2}+\phi_{\rm
                     disp}(d)\right]\,.
\end{equation}

Furthermore, for small separation ($d\rightarrow 0$) between the
two crystals with parallel optical axes, Eq.~(\ref{V2-maximum})
becomes

\begin{equation}\label{V-max-sm-parallel}
  V_{\rm p}(LD)={\rm sinc}\left(\frac{k_{\rm p}}{2d_{1}}|{\mathbf
M}L|^{2}\right)
\end{equation}

\noindent as opposed to the unity value predicted by the
single-mode theory. Note that this visibility is identical to that
of a single crystal of thickness $2L$ as was shown in previous
works \cite{ata01a}. Applying parameter values typical of our
experiments, we find that if $d_{1}=1$~m, $\lambda_{\rm
p}=351.1$~nm and $|{\bf M}|= 0.07$, then $V_{\rm p}\sim{\rm
sinc}(0.044\ L[{\rm mm}]^{2})$. For these conditions, therefore,
an observable deviation of $V_{\rm p}$ from unity can be realized
only for sufficiently large crystal thicknesses.

In the case of antiparallel optical axes, the visibility at the
center of the region of interference is

\begin{equation}\label{V-max-sm-antiparallel}
  V_{\rm a}(LD)=-1\,,
\end{equation}

\noindent which means that the coincidence-count rate is twice as
high at the center as it is on the shoulders, assuming the
analyzers are set so that $v_{\rm pol} = -1$ [see
Eq.~(\ref{R-final-2crys})]. In this case, the minus sign in
Eq.~(\ref{V-max-sm-antiparallel}) arises from a sign difference
between the quadratic susceptibilities of the two crystals, and
hence a sign difference between the spatial walkoff vectors
${\mathbf M}_1$ and ${\mathbf M}_2$. In effect, the spatial
walkoff in one crystal compensates for the spatial walkoff in the
other, and the spatial effects due to transverse wave vectors are
cancelled out. Exploitation of this effect is common in the design
of optical parametric oscillators.

\subsubsection{Multi-Parameter Formalism: Spatial Effects}

The quantum state generated from SPDC, which is concurrently
entangled in frequency and transverse wave vector, leads to
transverse spatial effects that can be observed in quantum
interference. As the aperture diameters are increased or the
apertures are brought closer to the output plane of the nonlinear
medium, a greater range of wave vectors is allowed to propagate
through to the detectors. These transverse wave vectors introduce
distinguishability between the signal and idler photons, thus
reducing the visibility of the observed quantum interference. This
is analogous to the temporal distinguishability introduced by the
use of a femtosecond-pulsed pump.

Equation~(\ref{V2-general}) is valid for any linear optical
system. However, to enable swift evaluation of the integrals in
this equation we approximate the circular apertures used in the
experiments by ``soft" Gaussian apertures of $(1/e)$ widths
$r_{\rm A}$ and $r_{\rm B}$. A sharp circular aperture, of the
type used in experiments, has a diffraction pattern described by a
first-order Bessel function, whereas a Gaussian aperture, of the
type used in the numerical simulations, has a Gaussian diffraction
pattern. Despite this fundamental difference, it is a fair
approximation if the width $r$ of the Gaussian is selected to
roughly fit the width $b$ of the Bessel function. In our
calculations, this is done by choosing $r = b/2\sqrt{2}$. This
approximation offers an indispensable advantage, as it allows us
to evaluate the ${\mathcal G}$-functions analytically (see
Appendix~\ref{sp-int}) and thereby reduce the demand for numerical
integration in making theoretical predictions. Under this
approximation, an expression for the visibility function at
$(\tau=LD)$ for parallel orientations of the optical axes is given
by [see Appendix~\ref{sp-int} and Eq.~(\ref{Spacial-integral})]

\begin{eqnarray}\label{V2-maximum-general-g-parallel}
  V_{{\rm p}}(LD)&=&2\frac{d_{1}(d_{1}+d)}{d_{1}^{2}+(d_{1}+d)^{2}}
                                     \frac{1}{\sqrt{1+\gamma^{2}}} \nonumber\\&&\times \,
                     \int^{1}_{0} d\zeta\
                     e^{-{\mathcal B}\left(1-\zeta\frac{d}{d+2d_{1}}\right)^{2}}
                     \cos\left[{\mathcal C}\zeta^{2}+{\mathcal D}\zeta +{\mathcal E}-\phi_{\gamma}(d)+\phi_{\rm disp}(d)\right]
\end{eqnarray}

\noindent where

\begin{equation}\label{gaussian-parameter}
  \gamma=\frac{k_{\rm p}r^{2}}{4d_{1}}\frac{d}{d_{1}+d}\,,\hspace{1cm}
\phi_{\gamma}(d)=-\arctan(\gamma)\,,
\end{equation}
and $r^2 = (r^2_{\rm A} + r^2_{\rm B})/2$. Here
\begin{eqnarray}\label{gaussian-parameter-integration}
  {\mathcal B}&=&2\left[\frac{k_{\rm p}|{\mathbf M}|Lr}{4(d_{1}+d)}\right]^{2}
                    \frac{1}{1+\gamma^{2}}\left(1+\frac{d}{2d_{1}}\right)^{2}\\
  {\mathcal C}&=&\frac{k_{\rm p}|{\mathbf M}L|^{2}}{8(d_{1}+d)}\frac{d}{d_{1}}
                    \frac{1}{1+\gamma^{2}}\\
  {\mathcal D}&=&-\frac{k_{\rm p}|{\mathbf M}L|^{2}}{2(d_{1}+d)}\left(1+\frac{d}{2d_{1}}\right)
                    \frac{1}{1+\gamma^{2}}\\
  {\mathcal E}&=&\frac{k_{\rm p}|{\mathbf M}L|^{2}}{8(d_{1}+d)}\left(\frac{d}{d_{1}}-4\gamma^{2}\frac{d_{1}+d}{d}\right)
                    \frac{1}{1+\gamma^{2}}.
\end{eqnarray}

\noindent Figure 7 presents a plot of the modulus of $V_{{\rm
p}}(LD)$ given in Eq.~(\ref{V2-maximum-general-g-parallel}) as a
function of the crystal separation $d$ for aperture diameters
$b=2.5$~mm, 4.0~mm, and 5.0~mm. Solid, shaded, and open circles
denote the maxima of visibility and correspond to 2.5-mm, 4.0-mm,
and 5.0-mm aperture diameters, respectively. For this plot,
$L=0.5$~mm, $\lambda_{\rm p} = 2\pi/k_{\rm p} = 351.1$~nm, and
$|M|= 0.07$ in keeping with our experiments.

For antiparallel axes, the visibility function at $\tau=LD$ is
\begin{eqnarray}\label{V2-maximum-general-g-antiparallel}
  V_{{\rm a}}(LD)&=&-2\frac{d_{1}(d_{1}+d)}{d_{1}^{2}+(d_{1}+d)^{2}}
                                     \frac{1}{\sqrt{1+\gamma^{2}}} \nonumber\\&&\times \,
                     \int^{1}_{0} d\zeta\
                     e^{-{\mathcal B}\left(1-\zeta\right)^{2}}
                     \cos\left[{\mathcal
                     E}(1-\zeta)^{2}-\phi_{\gamma}(d)+\phi_{\rm disp}(d)\right].
\end{eqnarray}

\noindent For the case of two crystals in contact ($d=0$) we have
${\mathcal B}=2(k_{\rm p}|{\mathbf M}|Lr/4d_{1})^{2}$, ${\mathcal
C}=0$, ${\mathcal D}=-k_{\rm p}|{\mathbf M}L|^{2}/2d_{1}$, and
${\mathcal E}=0$. From Eq.~(\ref{V2-maximum-general-g-parallel}),
we now observe that the visibility at $\tau = LD$ for parallel
optical axes is
\begin{equation}\label{V-max-g-parallel}
  V_{\rm p}(LD)=\exp \left[-2\left(\frac{k_{\rm p}|{\mathbf M}|L
r}{4d_{1}}\right)^{2}\right]\ {\rm sinc}\left(\frac{k_{\rm
p}}{2d_{1}}|{\mathbf M}L|^{2}\right)
\end{equation}
while for antiparallel axes
\begin{equation}\label{V-max-g-antiparallel}
  V_{\rm a}(LD)=-\sqrt{2\pi}\,\frac{d_{1}}{k_{\rm p}|{\mathbf M}|L
  r}\ {\rm erf}\left(\frac{k_{\rm p}|{\mathbf M}|L r}{2\sqrt{2}d_{1}}
                           \right).
\end{equation}

\noindent Note that these visibilities depend on the aperture
diameter, as well as the crystal thickness (as was seen in the
previous section), and are once more markedly different from
$V_{\rm a} = V_{\rm p} = 1$ as predicted by the single-mode
theory. Equations~(\ref{V-max-g-parallel}) and
(\ref{V-max-g-antiparallel}) are plotted in Fig.~8 as a function
of Gaussian aperture width $r$ for three different crystal
thicknesses $L=$ 0.5~mm, 1.5~mm, and 5.0~mm. As predicted, the
plot of the visibility obtained with a parallel orientation of the
crystal axes (solid) reduces faster than the visibility for an
antiparallel orientation (dashed).

\section{Experiment}

\subsection{Experimental Arrangement}

The experimental arrangement is illustrated in Fig.~9.  A 200-mW
cw Ar$^+$-ion laser operated at 351.1~nm served as the pump. This
highly monochromatic laser beam was passed sequentially through a
cascaded pair of $\beta$-barium-borate (BBO) crystals each with
thickness 0.5~mm. The thickness of the air gap between the
crystals was varied from $d=2$~mm to 100~mm. The crystals were
aligned to produce pairs of orthogonally polarized photons in
degenerate collinear type-II spontaneous parametric
down-conversion ($\omega^{0}_{\rm s} = \omega^{0}_{\rm i} =
\omega^{0}_{\rm p}/2$, where $\omega^{0}_{\rm s}$,
$\omega^{0}_{\rm i}$, and $\omega^{0}_{\rm p}$ represent the
central frequencies of the signal, idler, and pump fields,
respectively). The laser power was sufficiently low to ensure,
with high probability, that at most one photon pair was generated
at a given time. The high visibility obtained from separate
single-crystal quantum-interference experiments confirmed the
validity of this assumption. A dichroic mirror, which transmits
the 702-nm signal and idler beams while reflecting the 351-nm
pump, was placed after the two crystals to remove residual pump
laser beam.

The relative optical-path delay $\tau$ was introduced using a
$z$-cut birefringent crystalline-quartz plate of variable
thickness. The range of transverse wave vectors for the
down-converted light was selected by circular apertures of
diameters 2.5~mm to 5.0~mm. These apertures were positioned at
750~mm from the output plane of the second crystal. The single
aperture used in the experimental setup is formally equivalent to
the use of two apertures of identical diameters in the
interferometer of Fig.~4, as used for the theoretical discussions
of previous sections. The beams of down-converted light were then
directed to a nonpolarizing beam splitter, and thence to the two
arms of a polarization intensity interferometer. Each arm of the
interferometer comprised a Glan-Thompson polarization analyzer set
at $45^{\circ}$ with respect to the horizontal axis in the
laboratory frame, establishing the basis for the polarization
measurements. This basis was selected so as to permit observation
of the quantum-interference pattern as a function of the
optical-path delay $\tau$. Finally, a convex lens (not shown in
Fig.~9) was used to reduce the beam size to be less than the area
of the detector, an actively quenched Peltier-cooled
photon-counting avalanche photodiode. No spectral filters were
used in any of the experiments. Coincidence detection was
performed using a 3-nsec integration window and corrections for
accidental coincidences were not necessary.

\subsection{Experimental Results}

First, we report quantum interference from crystals oriented with
parallel and antiparallel optical axes in the small-aperture
approximation. We demonstrate that the visibility at $\tau=LD$
varies sinusoidally with crystal separation $d$. Second, we
investigate spatial effects on the visibility at $\tau=LD$ arising
from the acceptance of a broader range of transverse wave vectors
as the aperture is opened.

\subsubsection{Parallel and Antiparallel Optical Axes}

In this section we investigate the quantum interference patterns
from SPDC in two identical crystals with parallel and antiparallel
optical axes. The details of the experimental setup can be found
in the previous section. The signal and idler fields were selected
by a 2.5-mm circular aperture positioned 750~mm after the second
crystal.

In the lower portion of Fig.~10, we plot the visibility function
at the center of the interference region, $V(LD)$ as defined in
Eq.~(\ref{V2-maximum}), as a function of the separation $d$
between the two 0.5-mm-thick BBO crystals. Each data point on the
graph was obtained by calculating the ratio $(R_{0}-R_{\rm
int})/R_{0}$ where $R_{0}$ and $R_{\rm int}$ are the
coincidence-count rates on the shoulders ($\tau < 0$ and $\tau
> 2LD$) and at the center of the region of interference
($\tau=LD$), respectively. The shaded circles plot the visibility
at $\tau=LD$ as a function of crystal separation $d$ when the
optical axes of the two crystals are parallel. The open circles
correspond to the case of antiparallel optical axes. As
anticipated by Eqs.~(\ref{V2-maximum-general-g-parallel})
and~(\ref{V2-maximum-general-g-antiparallel}), the modulation in
visibility with crystal separation is sinusoidal. The $\pi$-phase
shift between these two cases arises from the change in sign of
$\chi^{(2)}$ as the relative orientation of the optical axes is
inverted. The theoretical curves (solid) are plots of
Eqs.~(\ref{V2-maximum-general-g-parallel})
and~(\ref{V2-maximum-general-g-antiparallel}) assuming a Gaussian
aperture of width $r$ (defined as the diameter $b$ of the circular
apertures divided by $2\sqrt{2}$, the width at $1/e$), set to
satisfy $2\sqrt{2}\,r=2.5$~mm. The agreement of the theoretical
curves with the experimental data demonstrates the validity of the
Gaussian-aperture approximation. The remaining experimental
parameters for this plot are $d_{1}= 750$~mm, $\lambda_{\rm
p}=2\pi/k_{\rm p}=351.1$~nm, and $|M|= 0.07$.

In the top left inset of Fig.~10, we show a representative example
of quantum-interference patterns in which the visibility function
$V_{\rm p}(LD)$ for crystals with parallel optical axes is either
minimum or maximum. In the pattern exhibiting a peak (minimum),
the crystals were 17.5~mm apart, while in the pattern exhibiting a
dip (maximum) the crystals were 37.5~mm apart. Each data point
presented on the visibility graph (lower inset) is extracted from
such interference patterns. In the figure at the top right, where
the two crystals are oriented with antiparallel optical axes, the
opposite is true: the pattern exhibiting a dip (maximum)
corresponds to a 17.5-mm crystal separation, while the pattern
exhibiting a peak (minimum) corresponds to a 37.5-mm separation.
Note from Eq.~(\ref{v-pol}) that when both polarization analyzers
are set to $45^{\circ}$, the polarization projection factor
$v_{\rm pol} = -1$. Thus, from Eq.~(\ref{R-final-2crys}), positive
values of $V(LD)$ correspond to a quantum interference dip while
negative values correspond to a quantum interference peak.
Orienting the crystals with antiparallel optical axes produces a
sign difference between the second-order nonlinearities of the two
crystals, so this condition is again reversed.

\subsubsection{Spatiotemporal Effects}

In previous work with single-crystal SPDC \cite{ata01a}, we found
that the acceptance of transverse wave vectors had substantial
effects on quantum-interference patterns. To investigate
spatiotemporal effects in dual-crystal SPDC, we carried out
identical experiments to those detailed in the previous section,
except that the 2.5-mm circular aperture was replaced with 4.0-mm
and 5.0-mm circular apertures. The apertures remained positioned
750~mm from the output plane of the second crystal.

As with the single-crystal experiments, we found that increasing
the diameter of the aperture lowers the visibility of the quantum
interference patterns. This effect can be explained by considering
the two independent mechanisms that introduce phases between the
two-photon probability amplitudes for each crystal: 1)~dispersion
in the air separating the two crystals, which results in the phase
$\phi_{\rm disp}$ [Eq.~(\ref{air-dispersion})], and 2)~the angular
spread of down-converted light which results in the phase
$\phi_\gamma$ [Eq.~(\ref{gaussian-parameter})]. Whereas $\phi_{\rm
disp}$ is dependent on the crystal separation $d$, but independent
of $r$ and $d_{1}$, $\phi_\gamma$ depends on all three parameters.
However, when $d/d_{1}\gg 1$, $\phi_{\gamma}\sim-\arctan(k_{\rm
p}r^{2}/4d_{1})$. In this limit, $\phi_\gamma$ is dependent on $r$
and $d_{1}$, but independent of the crystal separation $d$. As
such, $\phi_{\gamma}$ dominates when the aperture is sufficiently
large, while $\phi_{\rm disp}$ dominates when the aperture is
sufficiently small.

Fig.~11 displays plots of the visibility function at $\tau=LD$ as
a function of the crystal separation $d$ for 2.5-mm (top), 4.0-mm
(middle), and 5.0-mm (bottom) aperture diameters. The two crystals
are oriented with parallel axes. Note that for a fixed value of
$d$, there is a reduction of visibility for increased aperture
diameter. A reduction of visibility also occurs as the distance
between the crystals is increased. Moreover, the period of
oscillation contracts slightly as the aperture diameter is
increased.

\section{Conclusion}

We have developed a theory of Type-II spontaneous parametric
down-conversion (SPDC) in media with inhomogeneous distributions
of nonlinearity. The down-converted light can be concurrently
entangled in frequency, wave vector, and polarization. We have
shown that the state function of the down-converted light can be
controlled by design of the nonlinearity profile in the crystal,
as well as the spatial and spectral profile of the pump field. As
one rudimentary design, we have considered the case of two
nonlinear crystals separated by a linear dispersive medium. We
anticipate that even greater control can be identified and
exploited when other distributions of nonlinearity are employed.

The multiparameter formalism of SPDC in quantum
interferometry~\cite{ata01a} has been extended to incorporate the
longitudinal nonlinearity profile of the medium. The
quantum-interference pattern was shown to critically depend on the
specific design of the nonlinearity profile. We studied the case
of a cascaded pair of bulk crystals and experimentally verified
the theoretical predictions. We have demonstrated that the quantum
interference is sensitive to the medium between the crystals, as
well as the design of the optical system for the down-converted
light. In particular, collective interference effects were seen
between the two probability amplitudes corresponding to detection
of a photon pair generated in either crystal. The visibility of
this collective interference between the two amplitudes depends on
a relative phase, which is a function of the dispersion in the
linear medium and the acceptance angle of the optical system. In
the small-aperture approximation, we can continuously sweep the
quantum-interference pattern from a high-visibility triangular dip
to a high-visibility peak, simply by changing the distance between
the two crystals. In principle, the same effect can also be
observed when the dispersion properties of the linear medium are
changed for a fixed distance between the crystals. Furthermore, as
we increased the aperture diameter, and thus admitted a greater
range of transverse wave vectors into the optical system, we
observed a contraction in the oscillation period of the
visibility.

Our findings are expected to be of interest to the development of
SPDC sources using multiple-crystal configurations and/or
periodically poled materials, and to the advancement of quantum
technologies through quantum-state engineering.

{\em Acknowledgments.---}This work was supported by the National
Science Foundation, DARPA, and the David \& Lucile Packard
Foundation. The authors thank M.~C.~Booth for his invaluable help
with the manuscript.

\appendix

\section{The Visibility Function $V(\tau)$}\label{App-2}

The purpose of this Appendix is to derive the visibility function
$V(\tau)$ defined in Eq.~(\ref{V2-general}).  We consider the
special case of degenerate collinear Type-II SPDC from two
cascaded crystals of identical materials but different
thicknesses. This source of SPDC is used in the interferometer
shown in Fig.~4 with symmetric apertures and no spectral filters.
Given this explicit configuration, we evaluate
Eqs.~(\ref{R-Inf-r-2crys}), (\ref{R-0-r-2crys}),
(\ref{R-Inf-12-2crys}), and (\ref{R-0-12-2crys}) and rearrange the
results to obtain Eq.~(\ref{V2-general}).

To calculate ${\mathcal F}_{\rm AB}$ in Eq.~(\ref{A-F-D1}), we
need to consider the explicit form of the transfer function $H$
for a given optical system. In the Fresnel approximation (which is
well satisfied under the conditions of our experiments) the
transfer functions $H_{i}$ ($i={\rm A},{\rm B}$) in
Eq.~(\ref{A-F-D1}) are given by~\cite{ata01a}

\begin{eqnarray}\label{H-D-2crys}
    H_{i}({\mathbf x}_{i},{\mathbf q};\omega) & \propto &
        {\widetilde f}(\omega)\, e^{{\rm i}\frac{\omega}{c}(d_{1}+d_{2}+f)}\
        e^{-{\rm i}\frac{\omega}{2cf}|{\mathbf x}_{i}|^{2}\left(\frac{d_{2}}{f}-1 \right)}\
        e^{-{\rm i}\frac{cd_{1}}{2\omega}|{\mathbf q}|^{2}} \nonumber\\&&\hspace{1cm}\times \,
  \int\ d{\mathbf y}\ p_{i}({\mathbf y})\
        e^{-{\rm i}\left(\frac{\omega}{c f}{\mathbf x}_{i}-
        {\mathbf  q}\right)\cdot{\mathbf y}}
\end{eqnarray}

\noindent where ${\widetilde f}(\omega)$ is a (typically Gaussian)
spectral filter profile, $\omega = \omega^0_{\rm p} / 2 + \nu$ is
the frequency of the degenerate SPDC, $d_1$ is the distance from
the output plane of the second crystal to the aperture, $d_2$ is
the distance between the aperture and each detector, $f$ is the
focal length of both lenses, and $p_{i}({\mathbf y})$ is the pupil
function for aperture $i=({\rm A},{\rm B})$. In the absence of
interference filters [${\widetilde f}(\omega)=1$], the
quasi-monochromatic field approximation ($\nu \ll \omega_{\rm
p}^0$) allows the ${\mathcal F}_{\rm AB}$ functions in
Eq.~({\ref{A-F-D1}}) to become frequency-independent, as shown in
Eqs.~(\ref{FAB-appendix}) and (\ref{P-d-d1}).

The wave vector mismatch functions in Eq.~(\ref{zeta-r-simple})
and~(\ref{P-d-i}) are given under these approximations by

\begin{equation}\label{Delta-2crys}
  \Delta\left({\mathbf q},\nu\right)=- \nu D +
            \frac{2|{\mathbf q}|^{2}}{k_{\rm p}}+
            {\mathbf M}\cdot{\mathbf q}
\end{equation}

\noindent where $\Delta$ is the wave vector mismatch function due
to dispersion in the crystals, and $D$ and ${\mathbf M}$ are
material properties of the crystals \cite{Sasha-theory,rubin}.
Here $D=u^{-1}_{\rm o}-u^{-1}_{\rm e}$ is the dispersion
coefficient, where $u_{{\rm o},{\rm e}}$ are the group velocities
for ordinary and extraordinary waves at the central frequency
$\omega^{0}_{\rm p}/2$ and ${\mathbf q}=0$ (i.e. $\theta_{\rm
e}=\theta_{\rm p}$). In Eq.~(\ref{Delta-2crys}), ${\mathbf M} =
\frac{\partial}{\partial \theta_{\rm e}} \ln[n_{\rm
e}(\omega^{0}_{\rm p}/2,\theta_{\rm p})]{\mathbf e}_{\rm o.a.}$ is
the spatial walkoff vector, where ${\mathbf e}_{\rm o.a.}$ is a
unit vector pointing in the direction of the optical axis.

Meanwhile, the dominant contribution to the dispersion function
for the linear medium in Eq.~(\ref{P-d-i}) is

\begin{equation}\label{Delta-disp-air}
  \Delta^{\prime}=k_{\rm p}-{K_{\rm o}}-{K_{\rm e}}
\end{equation}

\noindent where $k_{\rm p}=n_{\rm l}(\omega^{0}_{\rm
p})\,\omega^{0}_{\rm p}/c$, $K_{\rm e}= n_{\rm d}(\omega_{\rm
e})\, \omega_{\rm e}/c$ and $K_{\rm o}= n_{\rm d}(\omega_{\rm o})
\,\omega_{\rm o}/c$ are the first-order expansions of $k_{\rm e}$
and $k_{\rm o}$, and $n_{\rm l}(\omega)$ is the index of
refraction in the linear medium. Evidently the dispersion in the
linear medium between the two crystals simply contributes a
difference in optical path length to the overall dispersion
function. For a delay line of thickness $l_{\tau}$, the phase
functions $\eta_{\tau}({\mathbf q},\nu)$ in Eq.~(\ref{delay-line})
are given under the Fresnel and quasi-monochromatic field
approximations by

\begin{equation}\label{delta-2crys}
  \eta_{\tau}({\mathbf q},\nu)=\left[-\nu D_{\tau}+
            \frac{2|{\mathbf q}|^{2}}{K_{\rm p}}+
            {\mathbf M}_{\tau}\cdot{\mathbf q} - K_{\rm o} - K_{\rm
            e}\right]\,l_{\tau}
\end{equation}

\noindent where $D_{\tau}$ and ${\mathbf M}_{\tau}$ are the
dispersion coefficient and the spatial walkoff vector for the
birefringent material of the delay line, respectively. If we
consider the contribution ${\rm e}^{-{\rm i}\eta_{\tau}({\bf
q},\,\nu)}$ for the delay line along with Equation
(\ref{zeta-r-simple}), we obtain

\begin{equation}\label{zeta-2crys}
  {\tilde \chi}^{(2)}_r({\mathbf q},\nu)\
  {\rm e}^{-{\rm i}\eta_{\tau}({\bf q},\,\nu) }=
        e^{-{\rm i}\Delta^{\prime} d\,\delta_{r,1}}\,
     \int^{0}_{-L_{r}}dz\
        {\mathcal Q}_{r}(\nu;z)\
        {\mathcal M}_{r}({\mathbf q};z)
\end{equation}

\noindent for $r = (1,2)$ where ${\tilde \chi}^{(2)}_1({\mathbf
q},\nu)$ and ${\tilde \chi}^{(2)}_2({\mathbf q},\nu)$ are the
nonlinearity profiles of the first and second crystals,
respectively. We have here introduced the quantities

\begin{eqnarray}\label{L-M-2crys}
   {\mathcal M}_{r}({\mathbf q};z)&=&
        e^{{\rm i}({\mathbf M}_{r}z-{\mbox{\small{\boldmath ${\mathcal L}$}}}_{r})\cdot{\mathbf q}}\
        e^{2{\rm i}\frac{z-\beta_{r}}{K_{\rm p}}|{\mathbf q}|^{2}}\\
   {\mathcal Q}_{r}(\nu;z) &=&
        e^{-{\rm i}\nu(D_{r}z+\tau_{r})}
\end{eqnarray}

\noindent where the parameters

\begin{eqnarray}
   \tau_{r}&=&\tau-L_{2}D\cdot\delta_{r,1}\label{tau-2crys}\\
   \beta_{r}&=&l_{\tau}+(L_{2}+d)\cdot\delta_{r,1}\label{Di-2crys}\\
   {\mbox{\small{\boldmath ${\mathcal L}$}}}_{r}&=&
                {\mathbf M}_{\tau}l_{\tau}+
                {\mathbf M}_{2}L_{3-r}\cdot\delta_{r,1}\label{M-2crys}
\end{eqnarray}

\noindent for $r=(1,2)$ and where $\tau=-l_{\tau}D_{\tau}$. Under
these conditions the single crystal components of the
coincidence-count rate are then given by
\begin{eqnarray}\label{R-Inf-r-simpl-2crys}
  R^{(r)}_{0}=\frac{\mu_{{\rm Ao},{\rm Be}}^{2}+\mu_{{\rm Bo},{\rm Ae}}^{2}}{D}\int dz\
                \Pi_{L_{r}}^{2}(z)\
                {\mathcal I}[-({\mathbf M}_{r}z+{\mbox{\small{\boldmath ${\mathcal L}$}}}_{r}),
                {\mathbf 0},s_{r},s_{r}]
\end{eqnarray} and
\begin{eqnarray}\label{R-0-r-simpl-2crys}
  R^{(r)}_{\rm int}&=&2\frac{\mu_{{\rm Ao},{\rm Be}}\ \mu_{{\rm Bo},{\rm Ae}}}{D}\int dz\
                \Pi_{L_{r}}(z)\
                \Pi_{L_{r}}\left(\frac{2\tau_{r}}{D}-z\right)\
                {\mathcal I}[-({\mathbf M}_{r}z+{\mbox{\small{\boldmath ${\mathcal L}$}}}_{r}),
                             {\mbox{\small{\boldmath ${\mathcal Z}$}}}_{r},s_{r},s_{r}]
\end{eqnarray}

\noindent where for brevity we introduce $\Pi_{L_{r}}(z)\equiv
{\rm rect}_{[0,L_{r}]}(z)$. Here the distance $s_r =
d_{1}-z+\beta_{r}$,

\begin{eqnarray}\label{Z-r-2crys}
  {\mbox{\small{\boldmath ${\mathcal Z}$}}}_{r}=
                    -2\left({\mathbf M}_{r}\frac{\tau_{r}}{D}+
                    {\mbox{\small{\boldmath ${\mathcal
                    L}$}}}_{r}\right),
\end{eqnarray}

\noindent and

\begin{eqnarray}\label{G-kn1}
    {\mathcal I}\left[{\mathbf z}_{0}(z),{\mathbf
    Z}(z,z^{\prime}),s_{k},s_{n}\right]&=&
    \int\ d{\mathbf q}\ d{\mathbf q^{\prime}}\, {\mathcal M}_{k}({\bf q};z) {\mathcal M}_{n}^{\ast}
    ({\bf q}^{\prime};z^{\prime})\ {\mathcal F}_{\rm AB}({\mathbf q},\pm{\mathbf
    q}^{\prime},\nu)\,.
\end{eqnarray}

We have assumed, in keeping with our experiments, apertures which
are symmetric such that $|p_{{\rm A},{\rm B}}({\mathbf
y})|=|p_{{\rm A},{\rm B}}({\mathbf -y})|$. Further details of
${\mathcal F}_{\rm AB}({\mathbf q},\pm{\mathbf q}^{\prime},\nu)$
can be found in Appendix~\ref{sp-int}. In experimental practice,
the distance $d_1$ between the output plane of the second crystal
and the aperture is much longer than the crystal thicknesses $L_1$
and $L_2$ and the optical path length $\l_\tau$ of the delay line.
As such, we may assume that

\begin{equation}\label{sr}
  s_{r}=d_{1}-z+\beta_{r}\sim d_{1}+(2-r)d.
\end{equation}

\noindent Also note that there are no approximations constraining
the value of $d$.

The baseline coincidence-count rate from collective interference
between the amplitudes from the two crystals is given by

\begin{eqnarray}\label{R-inf-simpl-2crys}
  R^{(12)}_{0}&=&\epsilon
        \frac{\mu_{{\rm Ao},{\rm Be}}^{2}+\mu_{{\rm Bo},{\rm Ae}}^{2}}{D}
                                        \nonumber\\&&\times \,
          \left\{\int dz\
                \Pi_{L_{1}}(z)\
                \Pi_{L_{2}}\left(z-\frac{\tau_{1}-\tau_{2}}{D}\right)\right.\
           {\mathcal I}[-({\mathbf M}_{1}z+{\mbox{\small{\boldmath ${\mathcal L}$}}}_{1}),
                             {\mbox{\small{\boldmath ${\mathcal Z}$}}}_{12}^{0}(z),s_{1},s_{2}]
                             \ e^{-{\rm i}\Delta^{\prime} d}+
                                        \nonumber\\
        &&+\int dz\
                \Pi_{L_{2}}(z)\ \left.
                \Pi_{L_{1}}\left(z-\frac{\tau_{2}-\tau_{1}}{D}\right)\
           {\mathcal I}[-({\mathbf M}_{2}z+{\mbox{\small{\boldmath ${\mathcal L}$}}}_{2}),
                             {\mbox{\small{\boldmath ${\mathcal Z}$}}}_{21}^{0}(z),s_{2},s_{1}]
                             \ e^{{\rm i}\Delta^{\prime}
                             d}\right\}
\end{eqnarray}

\noindent where

\begin{eqnarray}\label{Z-inf-2crys}
  {\mbox{\small{\boldmath ${\mathcal Z}$}}}_{12}^{0}(z)=
            -\left[z\left({\mathbf M}_{1}-{\mathbf M}_{2}\right)+
                        {\mathbf M}_{2}\frac{\tau_{1}-\tau_{2}}{D}+
                        {\mbox{\small{\boldmath
                        ${\mathcal L}$}}}_{1}-{\mbox{\small{\boldmath ${\mathcal
                        L}$}}}_{2}\right]
\end{eqnarray}

\noindent and the remaining parameters are given by

\begin{eqnarray}\label{para-null}
    \tau_{1}&=&\tau-L_{2}D \hspace{.5cm}
        s_{1}=d_{1}+d\hspace{.5cm}
            {\mbox{\small{\boldmath ${\mathcal L}$}}}_{1}={\mathbf M}_{\tau}l_{\tau}+{\mathbf M}_{2}L_{2}\\
    \tau_{2}&=&\tau \hspace{1.9cm}
        s_{2}=d_{1}\hspace{1.2cm}
            {\mbox{\small{\boldmath ${\mathcal L}$}}}_{2}={\mathbf
            M}_{\tau}l_{\tau}\,.
\end{eqnarray}

\noindent It is important to note that $R_{0}^{(12)} = 0$ for all
values of $\tau$, since in this case $\tau_{1}-\tau_{2}=-LD_{2}$
and the two rectangular functions $\Pi_{L_{r}}(z)$ in
Eq.~(\ref{R-inf-simpl-2crys}) never actually overlap. Thus the
coincidence-count rates are strictly constant outside the region
of interference. This lack of shoulder modulation is an important
indicator of the purity of the observed polarization-entangled
two-photon state. The collective interference term itself is given
by

\begin{eqnarray}\label{R-0-simpl-2crys}
  R^{(12)}_{\rm int}&=&2\epsilon
        \frac{\mu_{Ao,Be}\ \mu_{Bo,Ae}}{D}
                                        \nonumber\\&&\times \,
          \left\{\int dz\
                \Pi_{L_{1}}(z)\
                \Pi_{L_{2}}\left(\frac{\tau_{1}+\tau_{2}}{D}-z\right)\right.\
                {\mathcal I}[-({\mathbf M}_{1}z+{\mbox{\small{\boldmath ${\mathcal L}$}}}_{1}),
                             {\mbox{\small{\boldmath ${\mathcal Z}$}}}_{12}^{\rm int}(z),s_{1},s_{2}]
                             \ e^{-{\rm i}\Delta^{\prime} d}+ \nonumber\\
        &&+\int dz\
                \Pi_{L_{2}}(z)\ \left.
                \Pi_{L_{1}}\left(\frac{\tau_{2}+\tau_{1}}{D}-z\right)\
           {\mathcal I}[-({\mathbf M}_{2}z+{\mbox{\small{\boldmath ${\mathcal L}$}}}_{2}),
                             {\mbox{\small{\boldmath ${\mathcal Z}$}}}_{21}^{\rm int}(z),s_{2},s_{1}]
                             \ e^{{\rm i}\Delta^{\prime} d}\right\}
\end{eqnarray}

\noindent with

\begin{eqnarray}\label{Z-0-2crys}
  {\mbox{\small{\boldmath ${\mathcal Z}$}}}_{12}^{\rm int}(z)=
            -\left[z\left({\mathbf M}_{1}-{\mathbf M}_{2}\right)+
                        {\mathbf M}_{2}\frac{\tau_{1}+\tau_{2}}{D}+
                        {\mbox{\small{\boldmath ${\mathcal L}$}}}_{1}+{\mbox{\small{\boldmath ${\mathcal
                        L}$}}}_{2}\right]
\end{eqnarray}

\noindent where ${\mbox{\small{\boldmath ${\mathcal
Z}$}}}_{21}^{0,\rm int}$ can be found by interchanging all indices
$(1 \leftrightarrow 2)$ in ${\mbox{\small{\boldmath ${\mathcal
Z}$}}}_{12}^{0,\rm int}$.

With sufficient algebra, it is possible to arrange all of the
contributions to the coincidence rate in
Eqs.~(\ref{R-Inf-r-simpl-2crys}), (\ref{R-0-r-simpl-2crys}),
(\ref{R-inf-simpl-2crys}) and~(\ref{R-0-simpl-2crys}) to obtain
the structure of Eq.~(\ref{R-final-2crys}), where the baseline
coincidence-count rate is given by
\begin{equation}\label{Rinf-final-2crys}
  R_{0}=\left(\frac{k_{\rm p}}{2}\right)^{2}
                           \left[\frac{L_{1}}{Ds_{1}^{2}}+
                                 \frac{L_{2}}{Ds_{2}^{2}}\right]
                        \tilde{P}_{\rm A}({\bf 0})\
                        \tilde{P}_{\rm B}({\bf 0})\,,
\end{equation}

\noindent where $\tilde{P}_{i}({\mathbf q})$ is given by
Eq.~(\ref{P-d-d1}), and the polarization analyzer projection
factor is

\begin{equation}\label{v-pol-appendix}
  v_{\rm pol}=2\frac{\mu_{{\rm Ao},{\rm Be}}\ \mu_{{\rm Bo},{\rm Ae}}}{\mu_{{\rm Ao},{\rm Be}}^{2}+\mu_{{\rm Bo},{\rm
  Ae}}^{2}}\,.
\end{equation}

\noindent This, in turn, leads to Eq.~(\ref{V2-general}), the
desired visibility function in the special case where $L_1 = L_2 =
L$:

\begin{eqnarray}\label{V2-final-2crys}
  V(\tau)&=&\frac{1}{1+\rho^2} \int dz\
                \Pi_{L}(z)\
                \Pi_{L}\left(\frac{2\tau}{D}-2L-z\right)\
                     {\mathcal G}^{(1)}\left(\frac{z}{L};\frac{\tau}{LD}\right)\nonumber\\
              &&  +\frac{1}{1+\rho^2}\int dz\
                \Pi_{L}(z)\
                \Pi_{L}\left(\frac{2\tau}{D}-z\right)\
             {\mathcal
             G}^{(2)}\left(\frac{z}{L};\frac{\tau}{LD}\right)\nonumber\\
              &&  +2\epsilon \frac{\rho}{1+\rho^2}\int dz\
                \Pi_{L}(z)\
                \Pi_{L}\left(\frac{2\tau}{D}-L-z\right)\
                     \Re e\left[{\mathcal G}^{(12)}\left(\frac{z}{L}+\frac{1}{2};\frac{\tau}{LD}\right)
                                  \ e^{-{\rm i}\Delta^{\prime}
                                  d}\right]\,.
\end{eqnarray}

\noindent Here the functions ${\mathcal G}_{r=1,2}$ are given by

\begin{equation}\label{G-N-Functions}
   {\mathcal G}^{(r)}\left(\frac{z}{L};\frac{\tau}{LD}\right)=
   {\mathcal N}\left[-({\mathbf M}_{r}z+
            {\mbox{\small{\boldmath ${\mathcal L}$}}}_{r}),
                                  {\mbox{\small{\boldmath ${\mathcal Z}$}}}_{r},s_{r},s_{r}\right]
\end{equation}

\noindent for the single-crystal contributions, while the
component due to collective interference is given by

\begin{equation}
   {\mathcal G}^{(12)}\left(\frac{z}{L}+\frac{1}{2};\frac{\tau}{LD}\right)=
   {\mathcal N}\left[-({\mathbf M}_{1}z+{\mbox{\small{\boldmath ${\mathcal L}$}}}_{1}),
                                  {\mbox{\small{\boldmath ${\mathcal
                                  Z}$}}}_{12}^{0}(z),s_{1},s_{2}\right]\,.
\end{equation}

\noindent The function ${\mathcal N}$ is defined in
Appendix~\ref{sp-int} by Eq.~(\ref{Gen-norm-eval}), while the
normalization factor is given by $\rho=(d_{1}+d)/d_{1}$.

For a delay line comprised of a thin $z$-cut birefringent element
such as quartz, ${\mathbf M}_{\tau}={\mathbf 0}$ and
\begin{eqnarray}\label{l-null}
            {\mbox{\small{\boldmath
            ${\mathcal L}$}}}_{1}={\mathbf M}_{2}L_{2}\,,\hspace{2cm}
            {\mbox{\small{\boldmath ${\mathcal L}$}}}_{2}={\mathbf
            0}\,.
\end{eqnarray}

\noindent Under this assumption there are two important limits:
large separation between the crystals ($d\rightarrow \infty$) and
contact between the crystals ($d\rightarrow 0$). When the two
crystals are moved very far apart, we expect interference from
SPDC to be governed only by the second crystal, and the
quantum-interference pattern to be identical to that of SPDC from
a single crystal of thickness $L_2$. This can be seen from
Eq.~(\ref{V2-final-2crys}) by noting that in this case the
normalization factor $\rho \rightarrow \infty$ and only the last
of the three integrals in Eq.~(\ref{V2-final-2crys}) survives. In
this limit

\begin{eqnarray}\label{V2-final-2crys-approx1}
  V_{d\rightarrow \infty}(\tau)&=&
                \frac{1}{L}\int dz\
                \Pi_{L}(z)\
                \Pi_{L}\left(\frac{2\tau}{D}-z\right)\
                     {\mathcal G}^{(2)}\left(\frac{z}{L};\frac{\tau}{LD}\right)
\end{eqnarray}

\noindent and the shoulder normalization
factor~(\ref{Rinf-final-2crys}) is identical to that of a single
crystal of thickness $L$. When two crystals of the same material
with parallel optical axes are in contact, the quantum
interference pattern is identical to that obtained from a single
crystal of thickness $2L$. However, note that this does not hold
for antiparallel orientations of the optical axes of the two
crystals.

\subsection{Parallel Optical Axes}

If the crystals are oriented such that their optical axes are
parallel, ${\mathbf M}_{1}={\mathbf M}_{2}={\mathbf M}$ and the
${\mathcal G}$ functions in Eq.~(\ref{V2-general}) are specified
by
\begin{eqnarray}\label{G-parallel}
    {\mathcal G}_{\rm p}^{(1)}(\zeta;x)&=&{\mathcal N}\left[-{\mathbf M}L(\zeta+x),
                                  -2{\mathbf M} L x,d_{1}+d,d_{1}+d\right]\\
    {\mathcal G}_{\rm p}^{(12)}(\zeta;x)&=&{\mathcal N}\left[-{\mathbf M}L(\zeta+x),
                                    -2{\mathbf M}L x,d_{1}+d,d_{1}\right]\\
    {\mathcal G}_{\rm p}^{(2)}(\zeta;x)&=&{\mathcal N}\left[-{\mathbf M}L(\zeta+x),
                                  -2{\mathbf M} L
                                  x,d_{1},d_{1}\right]
\end{eqnarray}

\noindent where $\zeta = z / L$, $x=\tau / LD$, and ${\mathcal N}$
is defined in Eq.~(\ref{Gen-norm-eval}).

\subsection{Antiparallel Optical Axes}

The antiparallel orientation of the optical axes of the two
crystals gives ${\mathbf M}_{1}=-{\mathbf M}_{2}={\mathbf M}$ and
the ${\mathcal G}$-functions are specified by
\begin{eqnarray}\label{G-antiparallel}
    {\mathcal G}_{\rm a}^{(1)}(\zeta;x)&=&{\mathcal N}\left[-{\mathbf M}L(\zeta+x-2),
                                  -2{\mathbf M} L (x-1),d_{1}+d,d_{1}+d\right]\\
    {\mathcal G}_{\rm a}^{(12)}(\zeta;x)&=& {\mathcal N}\left[-{\mathbf M}L(\zeta+x-2),
                                    -2{\mathbf M}L(\zeta-1),d_{1}+d,d_{1}\right]\\
    {\mathcal G}_{\rm a}^{(2)}(\zeta;x)&=&{\mathcal N}\left[{\mathbf M}L(\zeta+x),
                                  2{\mathbf M} L
                                  x,d_{1},d_{1}\right].
\end{eqnarray}

\section{The Function ${\mathcal N}$}\label{sp-int}

In this Appendix, we derive an explicit form of the function
${\mathcal N}$, which appears in Eq.~(\ref{G-N-Functions}). We
define

\begin{equation}\label{N-I-defition}
   {\mathcal N}({\mathbf z}_{0},{\mathbf Z},s_{k},s_{n})=
                \frac{s_{k}s_{n}}{(k_{\rm p}/2)^{2}}
                \frac{{\mathcal I}({\mathbf z}_{0},{\mathbf Z},s_{k},s_{n})}{\tilde{P}_{\rm A}(0)
                         \tilde{P}_{\rm B}(0)}\nonumber
\end{equation}

\noindent where $\tilde{P}_{i}({\mathbf q})$, $k,n=(1,2)$, $s_1$
and $s_2$ are defined in Eqs.~(\ref{para-null}), and

\begin{equation}\label{I-defined}
    {\mathcal I}\left[{\mathbf z}_{0}(z),{\mathbf
    Z}(z,z^{\prime}),s_{k},s_{n}\right]=
    \int\ d{\mathbf q}\ d{\mathbf q^{\prime}}\, {\mathcal M}_{k}({\bf q};z) {\mathcal M}_{n}^{\ast}
    ({\bf q}^{\prime};z^{\prime})\ {\mathcal F}_{\rm AB}({\mathbf q},\pm{\mathbf
    q}^{\prime},\nu)\,,
\end{equation}

\noindent where ${\mathcal M}_{r}$ is defined in
Eq.~(\ref{L-M-2crys}). The function ${\mathcal F}_{\rm AB}$,
defined by Eq.~(\ref{A-F-D1}), under the Fresnel approximation and
in the absence of interference filters takes the form

\begin{equation}\label{FAB-appendix}
{\mathcal F}_{\rm AB}({\mathbf q},\pm {\mathbf q}^{\prime},\nu)=
   \tilde{P}_{\rm A}({\mathbf q}\mp{\mathbf q}^{\prime})\
   \tilde{P}_{\rm B}(-{\mathbf q}\pm{\mathbf q}^{\prime})\
              e^{-{\rm i}\frac{2d_{1}(\nu)}{k_{\rm p}}
              \left(|{\mathbf q}|^{2}-|{\mathbf
              q}^{\prime}|^{2}\right)}\,.
\end{equation}

\noindent Here

\begin{eqnarray}\label{P-d-d1}
  \tilde{P}_{i}({\mathbf q})&=&
        \int\ d{\mathbf y}\ p_{i}({\mathbf y})p^{\ast}_{i}({\mathbf y})\
            e^{-{\rm i}{\mathbf y}\cdot{\mathbf q}}\\
d_{1}(\nu)&=& d_1\left[1-\left(\frac{2\nu}{\omega^{0}_{\rm
p}}\right)^{2}\right]^{-1}
\end{eqnarray}

\noindent for $i = ({\rm A},{\rm B})$. Given these explicit forms,
Eq.~(\ref{I-defined}) becomes

\begin{equation}\label{Gen-eval}
   {\mathcal I}({\mathbf z}_{0},{\mathbf Z},s_{k},s_{n})=
          \frac{(k_{\rm p}/2)^{2}}{s_{k}s_{n}}\
          \exp \left[-{\rm i}\frac{k_{\rm p}}{8s_{n}}|{\mathbf Z}|^{2} \right]\
          \int\ d{\mathbf q}~
          \tilde{P}_{\rm A}(-{\mathbf q})\
          \tilde{P}_{\rm B}({\mathbf q})\
          {\mathcal W}\left({\mathbf q}+\frac{k_{\rm p}}{4s_{n}}{\mathbf Z}\right)\
          e^{-{\rm i}{\mathbf q}\cdot{\mathbf z}_{0}}
\end{equation}

\noindent where

\begin{equation}
 {\mathcal W}({\mathbf q})=-\frac{2{\rm i}d^{(kn)}_{r}}{\pi k_{\rm p}}
                          \exp\left[\frac{2{\rm i}d^{(kn)}_{r}}{k_{\rm p}}|{\mathbf
                          q}|^{2}\right]
\end{equation}

\noindent with

\begin{equation}
  \frac{1}{d^{(kn)}_{r}}=\frac{1}{s_{n}}-\frac{1}{s_{k}}.
\end{equation}

\noindent Finally, using Eq.~(\ref{Gen-eval}) we obtain

\begin{equation}\label{Gen-norm-eval}
   {\mathcal N}({\mathbf z}_{0},{\mathbf Z},s_{k},s_{n})=\exp \left[-{\rm i}\frac{k_{\rm p}}{8s_{n}}|{\mathbf Z}|^{2}
   \right]\ \int\ d{\mathbf q}\
          \tilde{{\mathcal P}}_{\rm A}(-{\mathbf q})\ \tilde{{\mathcal P}}_{\rm B}({\mathbf q})\
          {\mathcal W}\left({\mathbf q}+\frac{k_{\rm p}}{4s_{n}}{\mathbf Z}\right)
          e^{-{\rm i}{\mathbf q}\cdot{\mathbf z}_{0}}
\end{equation}

\noindent with $\tilde{{\mathcal P}}_{i}({\bf
q})=\tilde{P}_{i}({\bf q} )/\tilde{P}_{i}({\bf 0})$ for $i={\rm
A},{\rm B}$. In the limit $s_{k}\rightarrow s_{n}$,
Eq.~(\ref{Gen-norm-eval}) becomes

\begin{eqnarray}\label{Gen-eval-semp}
   {\mathcal N}({\mathbf z}_{0},{\mathbf Z},s_n,s_n)&=&
          \exp \left[-{\rm i}\frac{k_{\rm p}}{8s_n}|{\mathbf Z}|^{2}\right]\
          \tilde{{\mathcal P}}_{\rm A}\left(\frac{k_{\rm p}}{4s_n}{\mathbf Z}\right)\
          \tilde{{\mathcal P}}_{\rm B}\left(-\frac{k_{\rm p}}{4s_n}{\mathbf Z}\right)\
          e^{{\rm i}\frac{k_{\rm p}}{4s_n}{\mathbf Z}\cdot{\mathbf z}_{0}}\,.
\end{eqnarray}

\subsection{Small-Aperture Approximation}

Suppose that the apertures are so small that the pupil functions
$p_{{\rm A},{\rm B}}({\mathbf y})$ may be treated as delta
functions at ${\mathbf y} = 0$. In this case, $\tilde{{\mathcal
P}}_{\rm A}({\mathbf q})$ is a constant and
Eq.~(\ref{Gen-norm-eval}) becomes
\begin{eqnarray}\label{Gen-norm-eval-SM}
   {\mathcal N}_{\rm SM}({\mathbf z}_{0},{\mathbf Z},s_{k},s_{n})&=&
          \exp \left[-{\rm i}\frac{k_{\rm p}}{8}\left(\frac{|{\mathbf Z}-{\mathbf z}_{0}|^{2}}{s_{n}}-
                                            \frac{|{\mathbf
                                            z}_{0}|^{2}}{s_{k}}\right)\right].
\end{eqnarray}
In the limit $s_{k}\rightarrow s_{n}$ we then have
\begin{eqnarray}\label{Gen-norm-eval-SM-gamma0}
   {\mathcal N}_{\rm SM}({\mathbf z}_{0},{\mathbf Z},s_{n},s_{n})&=&
          \exp \left[-{\rm i}\frac{k_{\rm p}}{8s_{n}}\left(|{\mathbf Z}-{\mathbf z}_{0}|^{2}-
                                                 |{\mathbf
                                                 z}_{0}|^{2}\right)\right].
\end{eqnarray}

\subsection{Gaussian-Aperture Approximation}

To simplify the analysis, we consider the case of Gaussian
apertures with $p_{i}({\mathbf y})=\exp(-|{\mathbf
y}|^{2}/2r^{2}_{i})$ and $\tilde{{\mathcal P}}_{i}({\mathbf
q})=\exp(-|{\mathbf q}|^{2}r_{i}/4)$, so that
\begin{eqnarray}\label{Gen-norm-eval-G}
   {\mathcal N}_{\rm G}({\mathbf z}_{0},{\mathbf Z},s_{k},s_{n})&=&
          \frac{1-i\gamma}{1+\gamma^{2}}\
          \exp \left[{\rm i}\frac{k_{\rm p}}{8(s_{k}-s_{n})}|{\mathbf Z}|^{2}\right] \nonumber\\&&\times \,
          \exp\left[-{\rm i}\frac{k_{\rm p}}{8d^{(kn)}_{r}}
          \frac{1-i\gamma}{1+\gamma^{2}}
          \left|{\mathbf z}_{0}-\frac{s_{k}}{s_{k}-s_{n}}{\mathbf Z}\right|^{2}\right]
\end{eqnarray}

\noindent where

\begin{equation}\label{gamma}
  \gamma=\frac{k_{\rm p}}{4d^{(kn)}_{r}}\frac{r^{2}_{\rm A}+r^{2}_{\rm B}}{2}.
\end{equation}

\noindent It is useful to write the complex constant

\begin{equation}\label{gamma-costant}
  \frac{1-i\gamma}{1+\gamma^{2}}=\frac{1}{\sqrt{1+\gamma^{2}}}\
                                e^{-{\rm i}\arctan\gamma}
\end{equation}

\noindent so that we may write the phase

\begin{equation}\label{gamma-phase}
  \phi_{\gamma}=-\arctan\left(\frac{k_{\rm p}}{4d_{1}}\frac{r_{\rm A}^{2}+r_{\rm B}^{2}}{2}\frac{d/d_{1}}{1+d/d_{1}}\right).
\end{equation}

\noindent  In summary, the ${\mathcal N}$-function in the
Gaussian-aperture approximation is given by

\begin{eqnarray}\label{Gen-norm-eval-G2}
   {\mathcal N}_{\rm G}({\mathbf z}_{0},{\mathbf Z},s_{k},s_{n})&=&
          \frac{1}{\sqrt{1+\gamma^{2}}}\
          \exp \left[-\frac{k_{\rm p}}{8d^{(kn)}_{r}}
          \frac{\gamma}{1+\gamma^{2}}
          \left|{\mathbf z}_{0}-\frac{s_{k}}{s_{k}-s_{n}}{\mathbf Z}\right|^{2}\right] \nonumber\\&&\times \,
          \exp\left[-{\rm i}\frac{k_{\rm p}}{8d^{(kn)}_{r}}
          \frac{1}{1+\gamma^{2}}
          \left|{\mathbf z}_{0}-\frac{s_{k}}{s_{k}-s_{n}}{\mathbf Z}\right|^{2}\right] \nonumber\\&&\times \,
          \exp \left[{\rm i}\frac{k_{\rm p}}{8(s_{k}-s_{n})}|{\mathbf
          Z}|^{2}+{\rm i}\phi_{\gamma}\right].
\end{eqnarray}
In the limit $s_{k}\rightarrow s_{n}$, we have
\begin{eqnarray}\label{Gen-norm-eval-G2-gamma0}
   {\mathcal N}_{\rm G}({\mathbf z}_{0},{\mathbf Z},s_{n},s_{n})&=&
          \exp \left[-\left|\frac{k_{\rm p}{\mathbf Z}}{4s_{n}}\right|^{2}\frac{r^{2}_{\rm A}+r^{2}_{\rm B}}{4}\right]
                        \nonumber\\&&\times \,
          \exp\left[-{\rm i}\frac{k_{\rm p}}{8s_{n}}\left(|{\mathbf Z}-{\mathbf z}_{0}|^{2}-
                                                |{\mathbf
                                                z}_{0}|^{2}\right)\right].
\end{eqnarray}

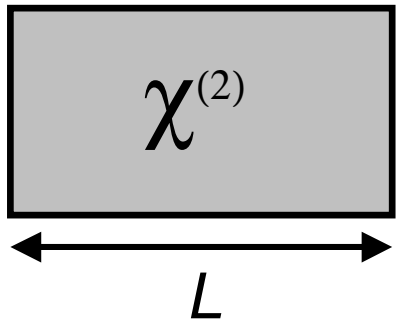
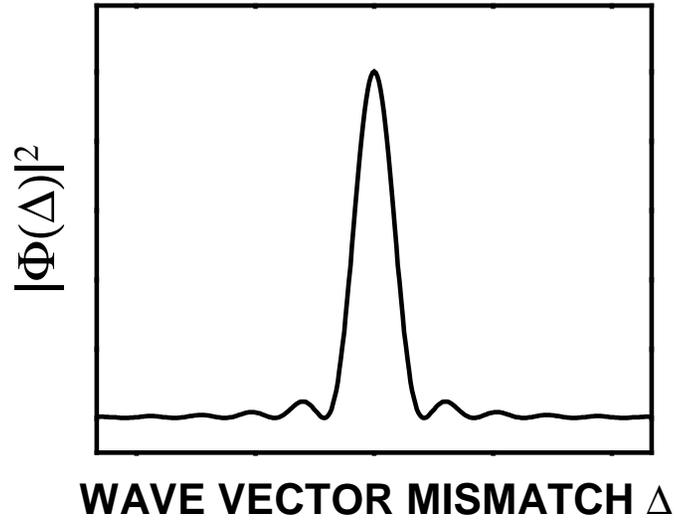
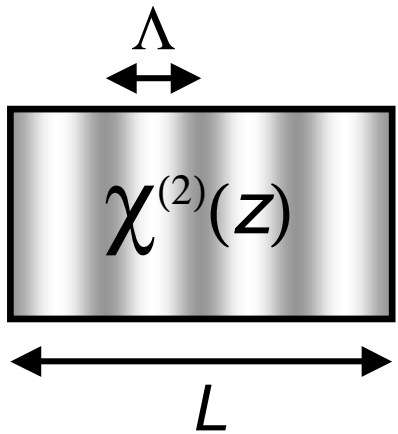
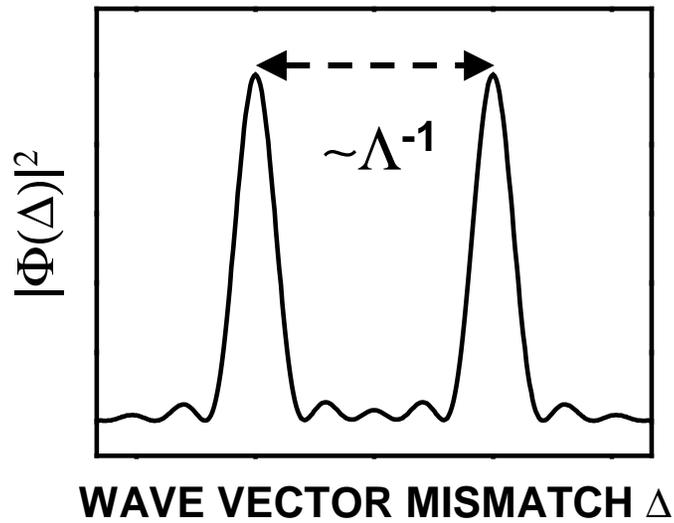
\begin{figure}
\label{figure-1} \caption{Control over the nonlinearity profile of
the generation medium allows control over the SPDC state function
$\Phi(\Delta)$. In the case of a monochromatic plane wave pump,
$\Phi(\Delta)$ is simply the inverse Fourier transform of the
nonlinearity profile $\chi^{(2)}(z)$. The upper figure shows a
sketch of $\left|\Phi(\Delta)\right|^2$ for Type-II SPDC from a
single bulk crystal of thickness $L$. The lower figure shows a
sketch of $\left|\Phi(\Delta)\right|^2$ for SPDC from a crystal
with sinusoidally varying nonlinearity. In principle, any
weighting profile of the signal and idler photons can be obtained
by a judicious choice of crystal structure.}
\end{figure}

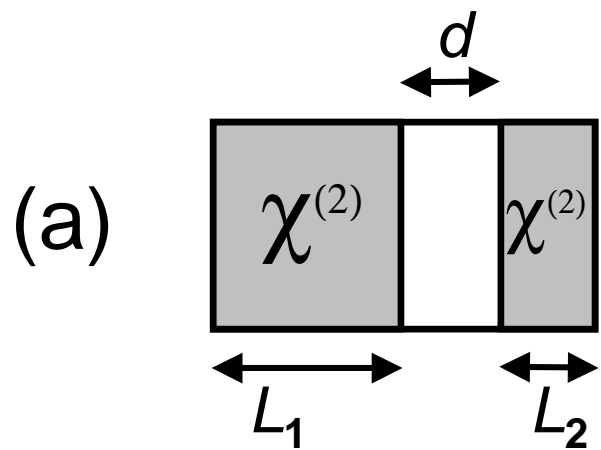
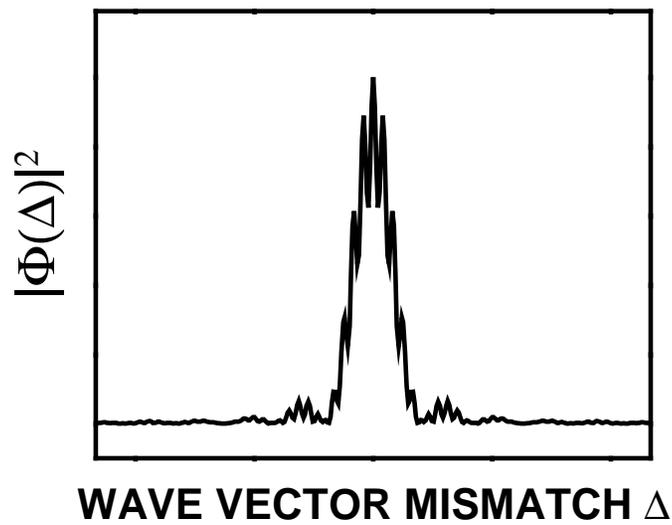
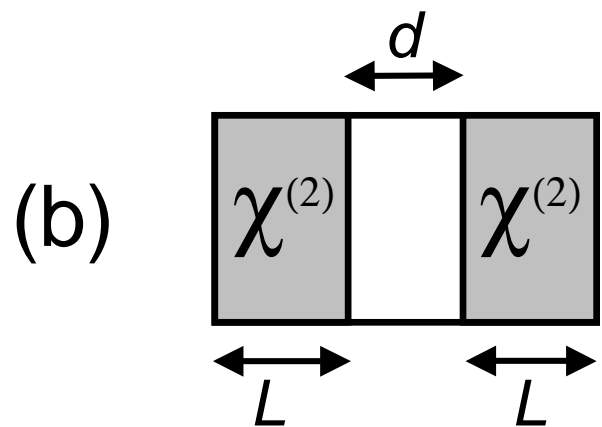
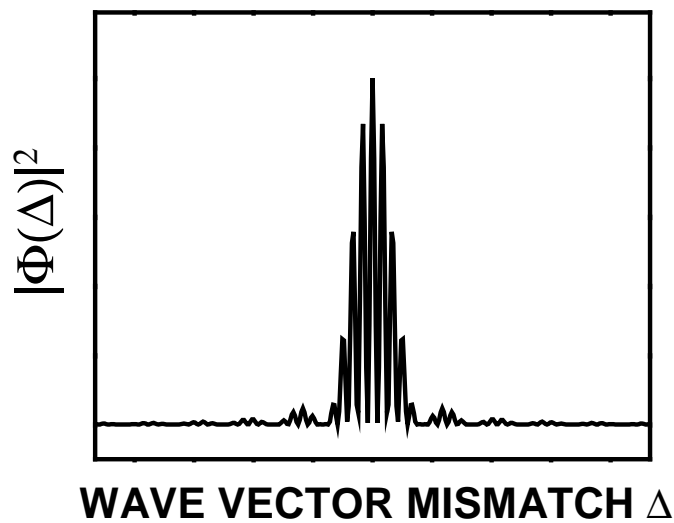
\begin{figure}
\label{figure-2} \caption{Impact of generation-medium symmetry on
the weighting profile of SPDC pumped by a monochromatic plane
wave. The upper figure shows $\left|\Phi(\Delta)\right|^2$ for
SPDC from two bulk crystals of unequal thickness separated by
dispersive linear medium. The lower figure shows
$\left|\Phi(\Delta)\right|^2$ for SPDC generated by a cascade of
two bulk crystals of equal thickness.}
\end{figure}

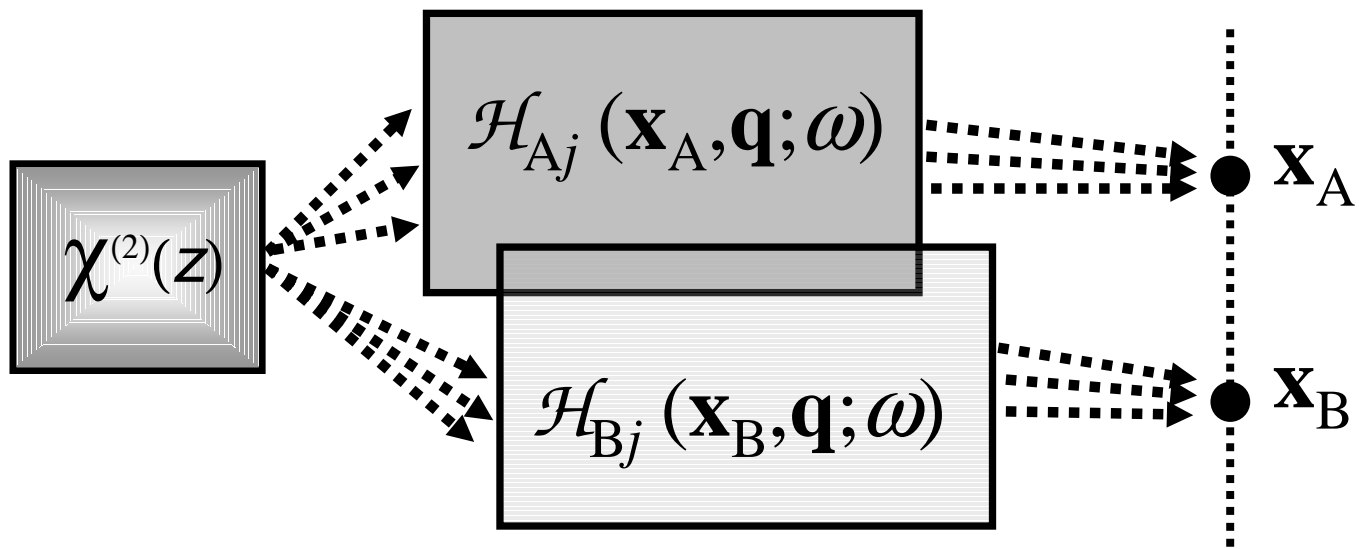
\begin{figure}
\label{figure-3} \caption{Propagation of the signal and idler
photons through arbitrary linear optical systems ${\mathcal
H}_{{\rm A}_j}$ and ${\mathcal H}_{{\rm B}_j}$.}
\end{figure}

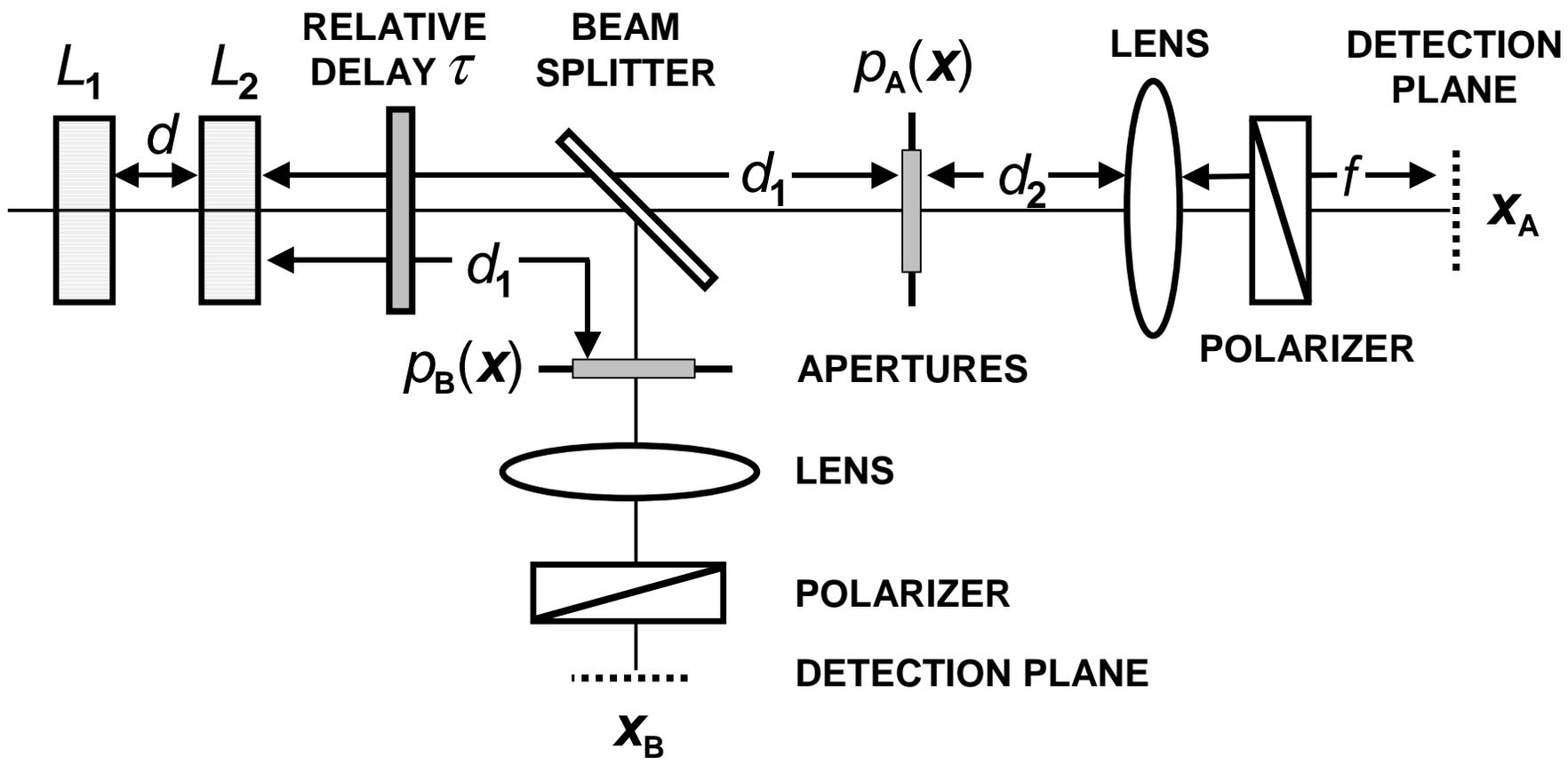
\begin{figure}
\label{figure-4} \caption{Schematic of a polarization
interferometer for which we compute quantum-interference patterns.
In this illustration, collinear SPDC is generated in two bulk
crystals of arbitrary thickness separated by an air gap.}
\end{figure}

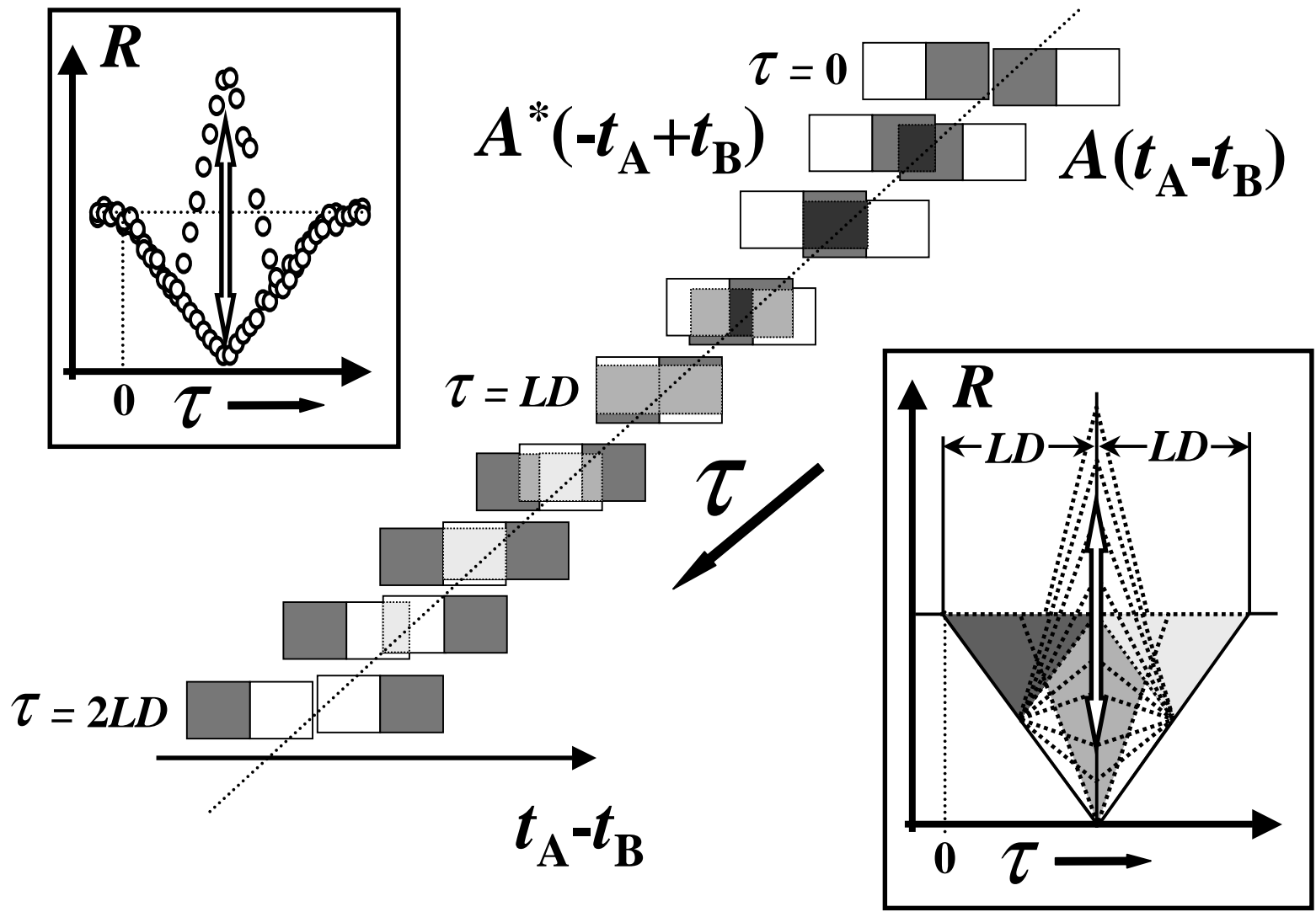
\begin{figure}
\label{figure-5} \caption{Sketch illustrating a heuristic approach
to calculating quantum interference in the single-mode limit for
two bulk crystals of the same material and thickness separated by
a linear dielectric. The results are coincidence rates as shown in
the bottom right inset. The coincidence-rate data in the upper
left inset is from representative experiments. Details are
provided in the text.}
\end{figure}

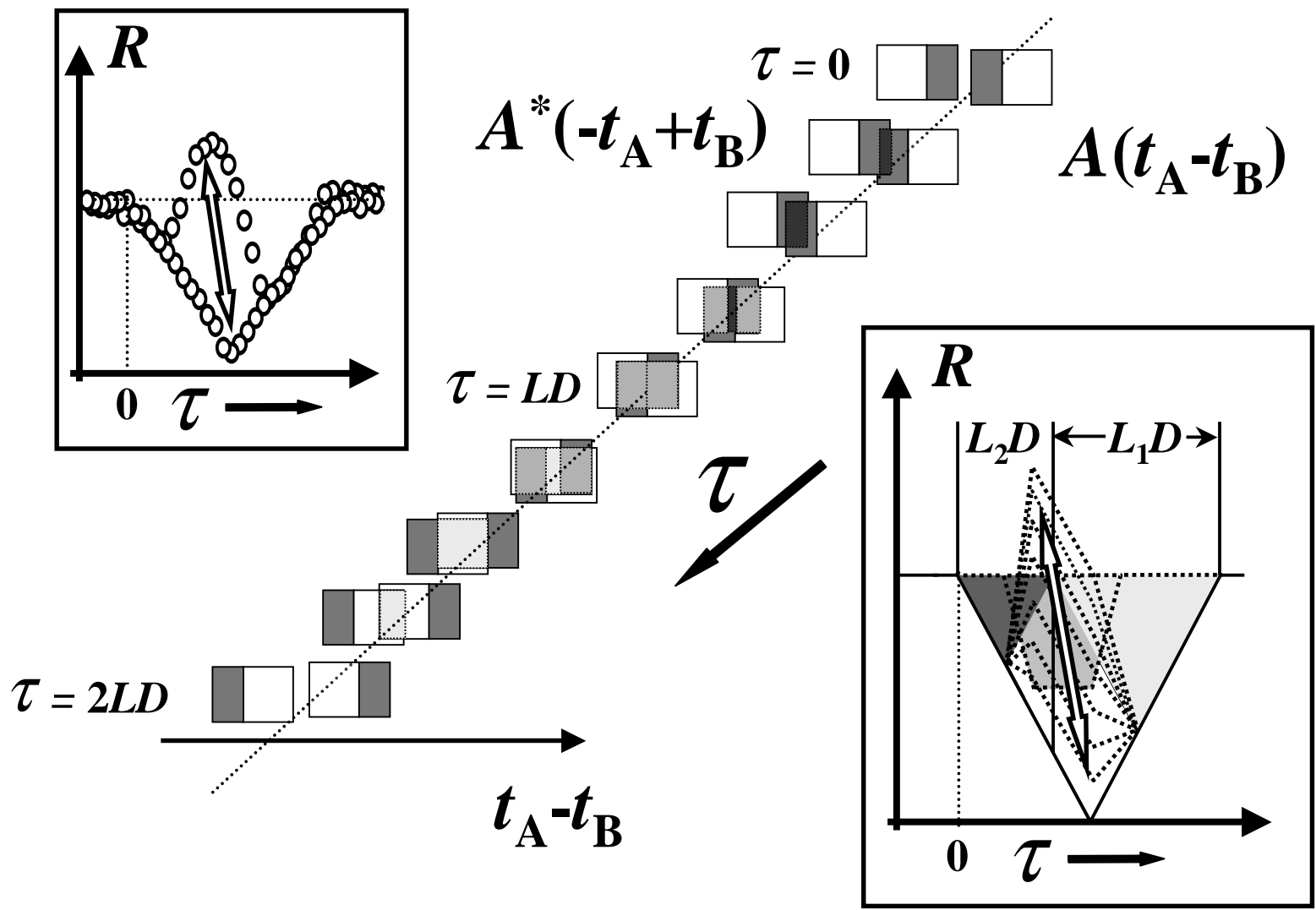
\begin{figure}
\label{figure-6} \caption{A figure analogous to that of Fig.~5 for
the case of two bulk crystals of unequal thickness.}
\end{figure}

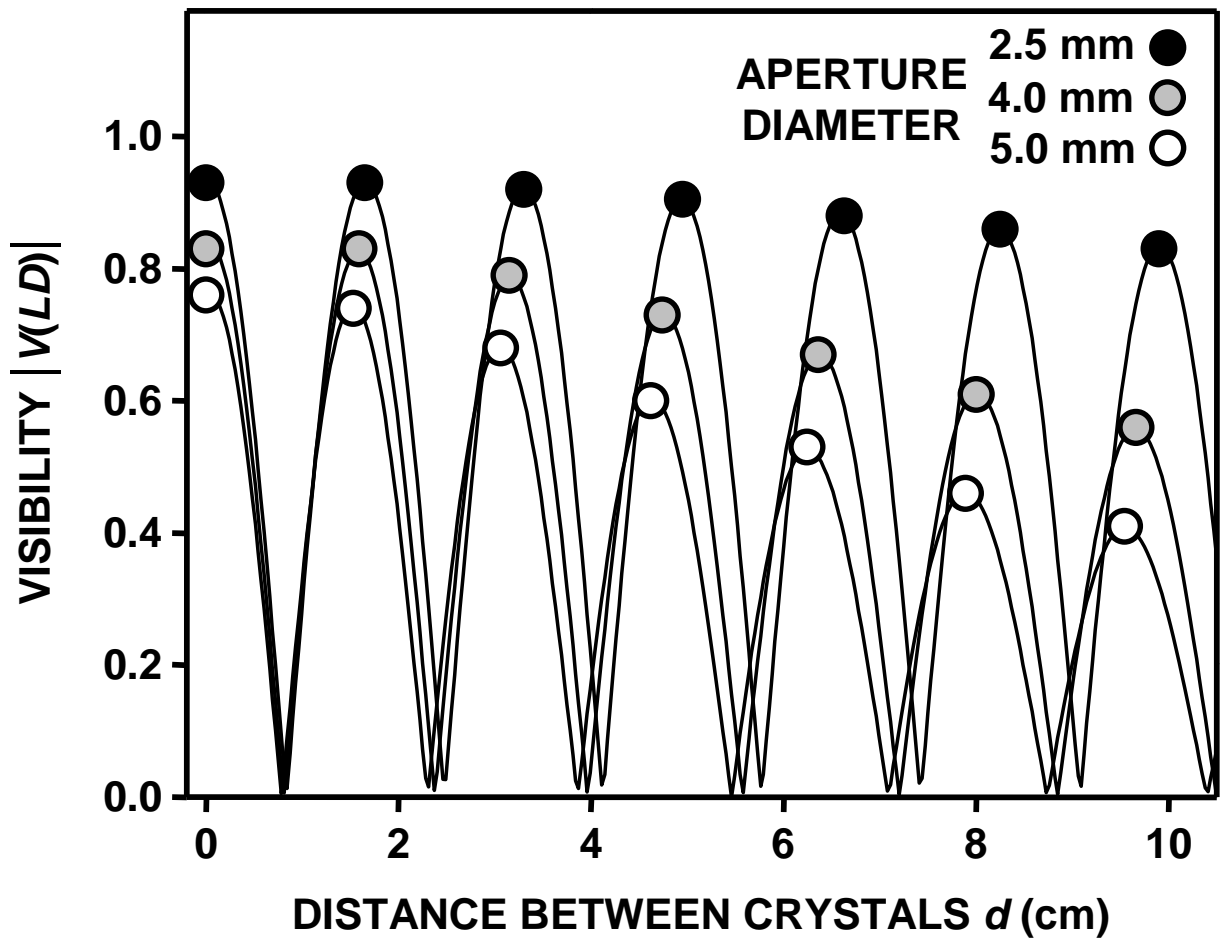
\begin{figure}
\label{figure-7} \caption{Visibility $\left|V(LD)\right|$ as a
function of crystal separation $d$ for the case of parallel
optical axes. The curves are plots of
Eq.~(\ref{V2-maximum-general-g-parallel}) for aperture diameters
$b=$ 2.5 mm, 4.0 mm, and 5.0 mm. Solid, shaded, and open circles
denote the maxima of visibility. The distance $d_1$ (see Fig.~4)
is 750 mm. Note that the period of visibility modulation as a
function of crystal separation $d$ contracts for increasing
aperture diameter.}
\end{figure}

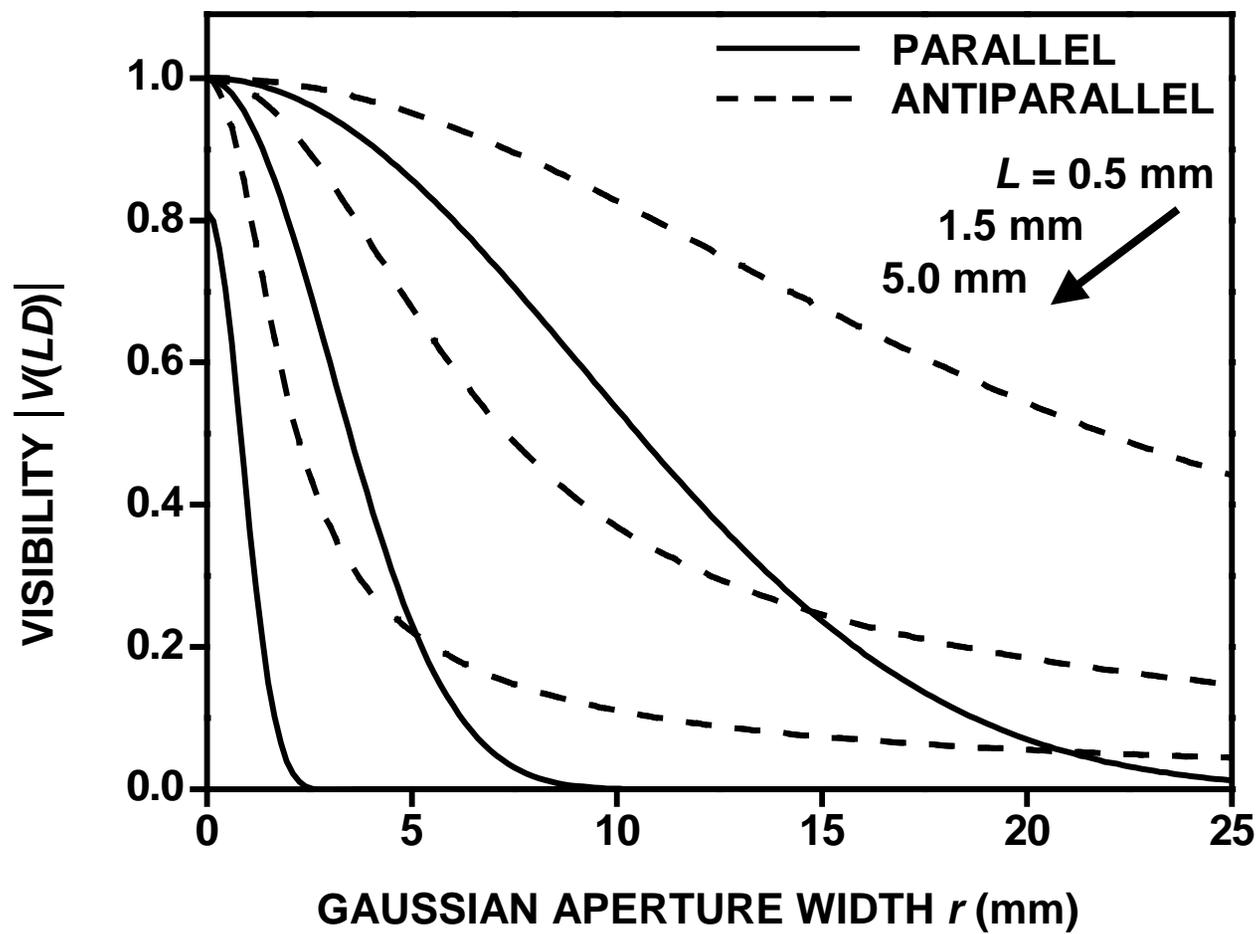
\begin{figure}
\label{figure-8} \caption{Visibility at $\tau = LD$ for two
identical crystals in contact ($d=0$) as a function of the $(1/e)$
width $r$ of identical Gaussian apertures in each arm of the
interferometer shown in Fig.~4. Solid (dotted) curves correspond
to parallel (antiparallel) optical axes. Results for different
crystal thicknesses $L$ are shown concurrently. The curves were
generated with the parameters $d_{1} = 1$~m, $\lambda_{\rm
p}=2\pi/k_{\rm p}=351.1$~nm and $|M| = 0.07$.}
\end{figure}

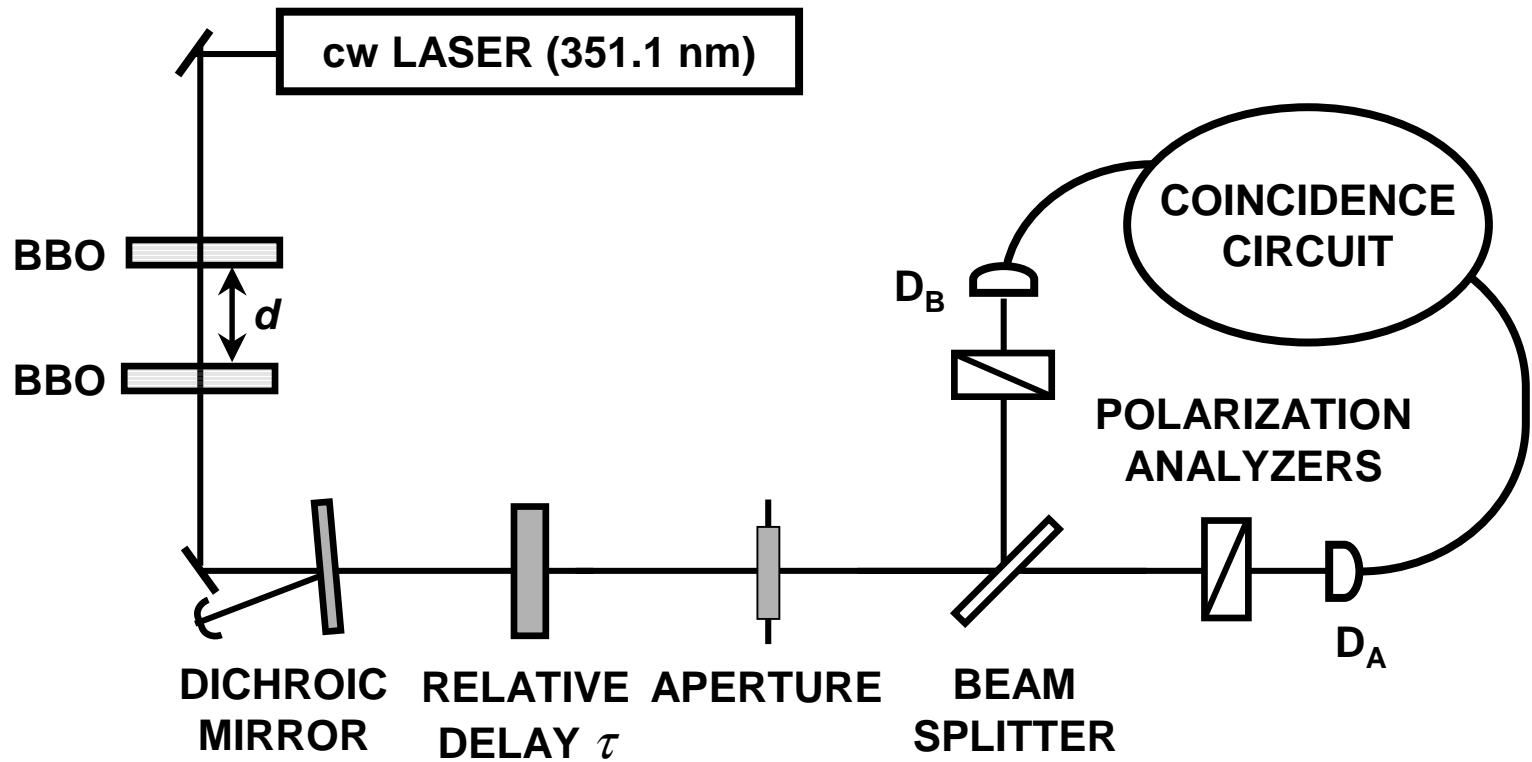
\begin{figure}
\label{figure-9} \caption{Experimental apparatus used to study
polarization interference in collinear SPDC from two cascaded
crystals separated by an air gap of thickness $d$. Details are
found in the text.}
\end{figure}

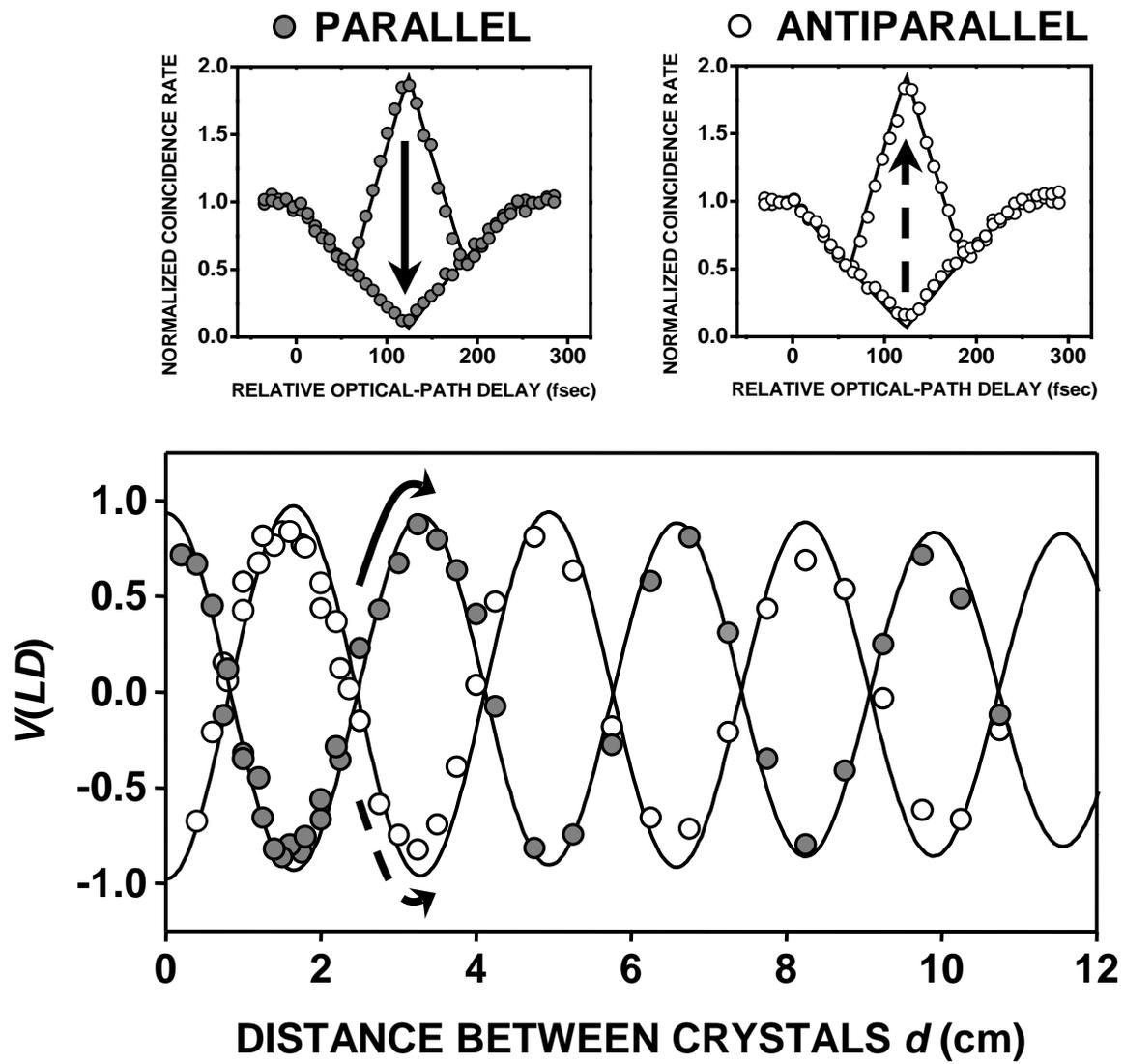
\begin{figure} \label{figure-10} \caption{
Visibility function at the center of the interference region,
$V(LD)$, as a function of crystal separation $d$. Symbols
represent data from parallel (shaded circles) and antiparallel
(open circles) optical axes. Note that in our experiments
$V(LD)=1$ corresponds to an interference dip, while $V(LD)=-1$
corresponds to an interference peak.  Insets at top are
representative interference patterns at extremes of $V(LD)$, taken
with $d=17.5$~mm and $d=32.5$~mm. Solid curves are plots of
Eqs.~(\ref{V2-maximum-general-g-parallel})
and~(\ref{V2-maximum-general-g-antiparallel}) for parallel and
antiparallel orientations, respectively.}
\end{figure}

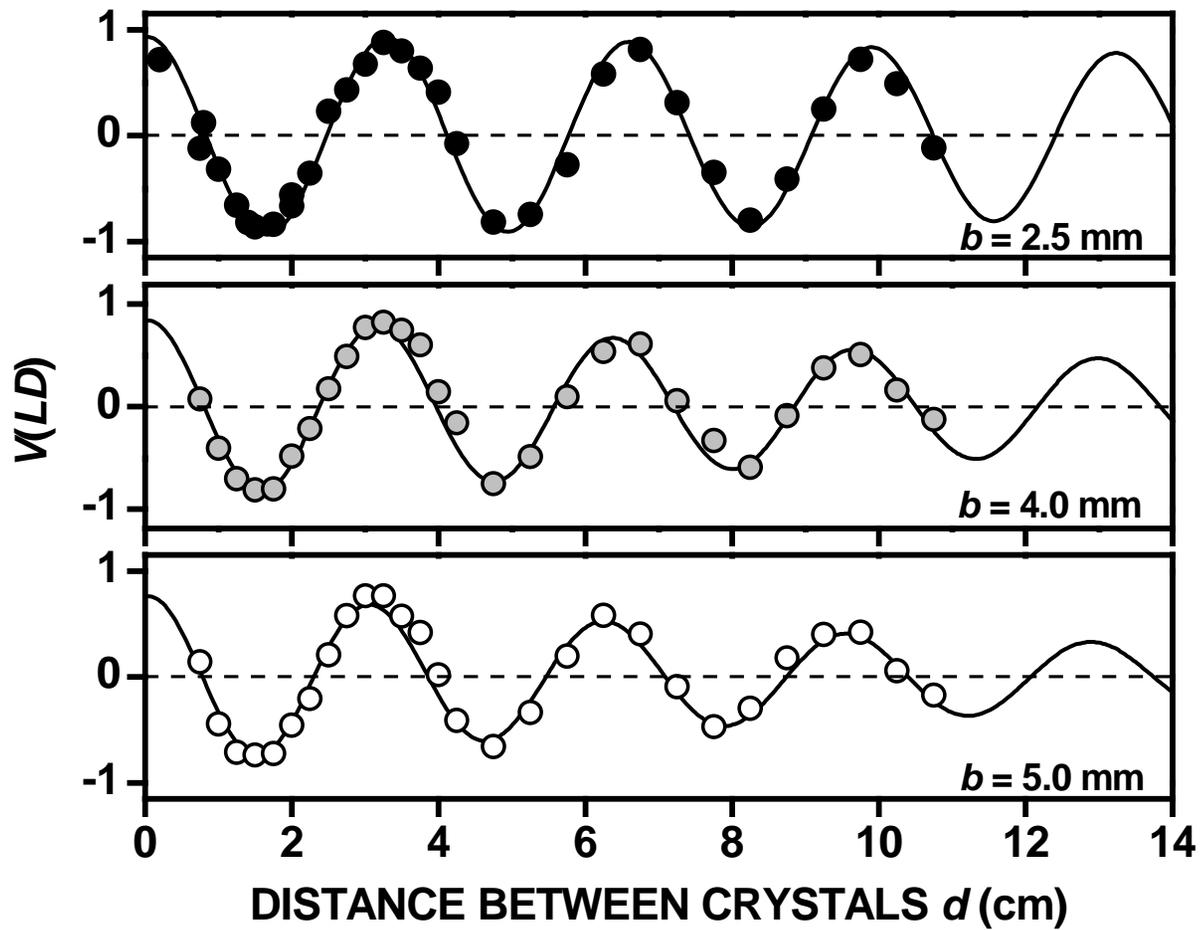
\begin{figure} \label{figure-11} \caption{
Visibility function $V(LD)$ as a function of crystal separation
$d$ for three different aperture diameters in the case of parallel
optical axes. Solid curves are plots of
Eq.~(\ref{V2-maximum-general-g-parallel}). Solid, shaded, and open
circles are experimental data for aperture diameters $b = $ 2.5
mm, 4.0 mm, and 5.0 mm, respectively. The distance $d_1$ (see
Fig.~4) is 750 mm.}
\end{figure}


\newpage


\begin{references}

\bibitem{SPDC} S.~E.~Harris, M.~K.~Oshman, and R.~L.~Byer, Phys.\ Rev.\ Lett.\ {\bf 18,} 732 (1967); D.~Magde and H.~Mahr, Phys.\ Rev.\ Lett.\ {\bf 18,} 905 (1967); D.~N.~Klyshko, {\it Photons and Nonlinear Optics} (Nauka,
Moscow, 1980); J.~Pe\v rina, Z.~Hradil, and B.~Jur\v co, {\it
Quantum Optics and Fundamentals of Physics} (Kluwer, Boston,
1994); L.~Mandel and E.~Wolf, {\it Optical Coherence and Quantum
Optics} (Cambridge, New York, 1995), Ch.\ 22.

\bibitem{Bell} J.~S.~Bell,~Physics {\bf 1}, 195 (1964); J.~F.~Clauser,
  M.~A.~Horne, A.~Shimony, and R.~A.~Holt, Phys.\ Rev.\ Lett.\ {\bf 23,} 880
  (1969); P.~G.~Kwiat, K.~Mattle, H.~Weinfurter, A.~Zeilinger,
  A.~V.~Sergienko, and Y.~H.~Shih, Phys.\ Rev.\ Lett.\ {\bf 75,} 4337 (1995).

\bibitem{Hardy} L.~Hardy, Phys.\ Rev.\ Lett.\ {\bf 71,} 1665
(1993); J.~R.~Torgesson, D.~Branning, C.H.~Monken, and L.~Mandel,
Phys.~Lett.~A, {\bf 204,} 323 (1995); G.~Di Giuseppe,
F.~De~Martini, and D.~Boschi, Phys.~Rev.~A {\bf 56,} 176 (1997);
D.~Boschi, S.~Branca, F.~De~Martini, and L.~Hardy,
Phys.~Rev.~Lett. {\bf 79,} 2755 (1997); A.~G.~White,
D.~F.~V.~James, P.~H.~Eberhard, and P.~G.~Kwiat, Phys.\ Rev.\
Lett.\ {\bf 83,} 3103 (1999).

\bibitem{GHZ} D.~M.~Greenberger, M.~A.~Horne, and A.~Zeilinger, in {\em
Bell's Theorem, Quantum Theory, and Conceptions of the Universe},
edited by M.~Kafatos (Kluwer, Dordrecht, 1989); D.~M.~Greenberger,
M.~A.~Horne, A.~Shimony, and A.~Zeilinger, Am.\ J.\ Phys.\ {\bf
58,} 1131 (1990); D.~Bouwmeester, J.-W.~Pan, M.~Daniell,
H.~Weinfurter, and A.~Zeilinger, Phys.\ Rev.\ Lett.\ {\bf 82,}
1345 (1999).

\bibitem{Metrology} D.~N.~Klyshko, Sov.\ J.\ Quantum Electron.\ {\bf
7,} 591 (1977); A.~Migdall, R.~Datla, A.~V.~Sergienko,
J.~S.~Orszak, and Y.~H.~Shih, Appl.\ Opt.\ {\bf 37,} 3455 (1998).

\bibitem{Fei-PRL} B.~E.~A.~Saleh, B.~M.~Jost, H.-B.~Fei, and M.~C.~Teich,
  Phys.\ Rev.\ Lett.\ {\bf 80,} 3483 (1998); H.-B.~Fei, B.~M.~Jost,
  S.~Popescu, B.~E.~A.~Saleh, and M.~C.~Teich, Phys.\ Rev.\ Lett.\ {\bf 78,}
  1679 (1997).

\bibitem{optical-imaging} B.~M.~Jost, A.~V.~Sergienko, A.~F.~Abouraddy,
  B.~E.~A.~Saleh, and M.~C.~Teich, Opt.\ Express {\bf 3,} 81 (1998);
  B.~E.~A.~Saleh, A.~F.~Abouraddy, A.~V.~Sergienko, and M.~C.~Teich, Phys.\
  Rev.\ A {\bf 62,} 043816 (2000); A.~F.~Abouraddy, B.~E.~A.~Saleh, A.~V.~Sergienko, and M.~C.~Teich, Phys.\
  Rev.\ Lett.\ {\bf 87,} 123602-1 (2001); A.~F.~Abouraddy, B.~E.~A.~Saleh, A.~V.~Sergienko,
  and M.~C.~Teich, Opt.\ Express {\bf 9,} 498 (2001); A.~F.~Abouraddy, B.~E.~A.~Saleh, A.~V.~Sergienko, and
  M.~C.~Teich, J.\ Opt.\ Soc.\ Am.\ B\ {\bf 19,} in press (2002); L.~A.~Lugiato, A.~Gatti,
and E.~Brambilla, J.\ Opt.\ B, in press (2002).

\bibitem{Cryptography} A.~K.~Ekert, Phys.\ Rev.\ Lett.\ {\bf 67,} 661 (1991);
J.~G.~Rarity and P.~R.~Tapster, Phys.\ Rev.\ A {\bf 45,} 2052
(1992); J.~Brendel, N.~Gisin, W.~Tittel, and H.~Zbinden, Phys.\
Rev.\ Lett.\ {\bf 82,} 2594 (1999); A.~V.~Sergienko,
M.~Atat\"ure,~Z.~Walton, G.~Jaeger, B.~E.~A.~Saleh, and
M.~C.~Teich, Phys.\ Rev.\ A {\bf 60,} R2622 (1999).

\bibitem{Teleportation} C.~H.~Bennett, G.~Brassard, C.~Crepeau, R.~Jozsa,
A.~Peres, and W.~K.~Wootters, Phys.\ Rev.\ Lett.  {\bf 70,} 1895
(1993); D.~Boschi, S.~Branca, F.~De~Martini, L.~Hardy, and
S.~Popescu, Phys.\ Rev.\ Lett.\ {\bf 80,} 1121 (1998);
D.~Bouwmeester, J.-W.~Pan, K.~Mattle, M.~Eibl, H.~Weinfurter, and
A.~Zeilinger, Nature {\bf 390,} 575 (1997).

\bibitem{Fs-SPDC} G.~Di~Giuseppe, L.~Haiberger, F.~De~Martini, and
A.~V.~Sergienko, Phys.\ Rev.\ A {\bf 56,} R21 (1997); W.~P.~Grice,
R.~Erdmann, I.~A.~Walmsley, and D.~Branning, Phys.\ Rev.\ A {\bf
57,} R2289 (1998); W.~P.~Grice and I.~A.~Walmsley, Phys.\ Rev.\ A
{\bf 56,} 1627 (1997); T. E.~Keller and M.~H.~Rubin, Phys.\ Rev.\
A {\bf 56,} 1534 (1997); J.~Pe\v {r}ina,~Jr., A.~V.~Sergienko,
B.~M.~Jost, B.~E.~A.~Saleh, and M.~C.~Teich, Phys.\ Rev.\ A {\bf
59,} 2359 (1999); D.~Branning, W.~P.~Grice, R.~Erdmann, and
I.~A.~Walmsley, Phys.\ Rev.\ Lett.\ {\bf 83,} 955 (1999);
T.~Tsegaye, J.~S\"{o}derholm, M.~Atat\"{u}re, A.~Trifonov,
G.~Bj\"{o}rk, A.~V.~Sergienko, B.~E.~A.~Saleh, and M.~C.~Teich,
Phys.\ Rev.\ Lett.\ {\bf 85,} 5013 (2000).

\bibitem{QIL-fs-SPDC} M.~Atat\"ure, A.~V.~Sergienko, B.~M.~Jost,
B.~E.~A.~Saleh, and M.~C.~Teich, Phys.\ Rev.\ Lett.\ {\bf 83,}
1323 (1999); M.~Atat\"ure, A.~V.~Sergienko, B.~E.~A.~Saleh, and
M.~C.~Teich, Phys.\ Rev.\ Lett.\ {\bf 84,} 618 (2000).

\bibitem{ata01a} M. Atat\"{u}re, G. Di Giuseppe, M. D. Shaw, A. V. Sergienko, B. E. A. Saleh, and
  M. C. Teich, Phys.\ Rev.\ A, in press (February 2002); M. Atat\"{u}re, G. Di Giuseppe, M. D. Shaw, A. V. Sergienko, B. E. A. Saleh, and
  M. C. Teich, submitted to Phys.~Rev.~A.

\bibitem{mandel} X.~Y.~Zhou, L.~J.~Wang, and L.~Mandel, Phys.\ Rev.\
  Lett.\ {\bf 67,} 318 (1991).

\bibitem{bur97} D. N. Klyshko, Zh. Eksp. Teor. Fiz. {\bf 104}, 2676 (1993) [JETP {\bf 77}, 222 (1993)];
                D. N. Klyshko, Laser~Physics {\bf 4}, 663 (1994);
                A. V. Burlakov, M. V. Chekhova, D. N. Klyshko, S. P. Kulik, A. N. Penin, Y. H. Shih and D. V. Strekalov,
                        Phys.~Rev.~A {\bf 56}, 3214 (1997);
                A. V. Burlakov, D. N. Klyshko, S. P. Kulik, A. N. Penin and M. V. Chekhova,
                        Pis'ma~Zh.~Eksp.~Teor.~Fiz. {\bf 65}, 20 (1997) [JETP~Lett. {\bf 65}, 19 (1997)].

\bibitem{zeilinger} T. J. Herzog, J. G. Rarity, H. Weinfurter, and
A. Zeilinger, Phys.\ Rev.\ Lett.\ {\bf 72,} 629 (1994); T. J.
Herzog, P.G. Kwiat, H. Weinfurter, and A. Zeilinger, Phys.\ Rev.\
Lett.\ {\bf 75,} 3024 (1995); P. W. Milonni, H. Fearn, and A.
Zeilinger, Phys.\ Rev.\ A.\ {\bf 53,} 4556 (1996).

\bibitem{grice} J.W. Pan, D. Bouwmeester, H. Weinfurter, and A.
Zeilinger, Phys.\ Rev.\ Lett.\ {\bf 80,} 3891 (1998); D.
Bouwmeester, J.-W. Pan, M. Daniell, H. Weinfurter, and A.
Zeilinger, Phys.\ Rev.\ Lett.\ {\bf 82,} 1345 (1999); D. Branning,
W. Grice, R. Erdmann, and I. A. Walmsley, Phys.\ Rev.\ A\ {\bf62,}
013814 (2000).

\bibitem{kwiat} P.~G.~Kwiat,  E.~Waks, A.~G.~White, I.~Appelbaum, and P.~H.~Eberhard, Phys.\ Rev.\ A\
  {\bf60,} R773 (1999); Y. H. Kim, S. P. Kulik, and Y. Shih,  Phys.\ Rev.\ A\  {\bf62,} 011802 (2000).

\bibitem{ata01} M. Atat\"{u}re, A. V. Sergienko, B. E. A. Saleh, and
  M. C. Teich, Phys.\ Rev.\ Lett.\ {\bf 86}, 4013 (2001).

\bibitem{synthesis} A.~G.~White, D.~F.~V.~James, W.~J.~Munro, and
P.~G.~Kwiat, Phys.\ Rev.\ A\ {\bf 65}, 012301 (2002).

\bibitem{fejer-PPLN}
N.~Bloembergen, U.S. Patent 3,384,443, May 21, 1968; M.~M.~Fejer,
G.~A.~Magel, D.~H.~Jundt, and R.~L.~Byer, IEEE
  J.\ Quant.\ Electron.\ {\bf 11}, 2631 (1992); P.~Baldi, P.~Aschieri, S.~Nouh,
  M.~De~Micheli, D.~B.~Ostrowsky, D.~Delacourt, and M.~Papuchon, IEEE
  J.\ Quant.\ Electron.\ {\bf 31}, 997 (1995); M.~A.~Arbore,
  A.~Galvanauskas, D.~Harter, M.~H.~Chao, and M.~M.~Fejer, Opt.\
  Lett.\, {\bf 22}, 1341 (1997); G.~Imeshev, A.~Galvanauskas,
  D.~Harter, M.~A.~Arbore, M.~Proctor, and M.~M.~Fejer, Opt.\
  Lett.\ {\bf 23}, 864 (1998); M.~H.~Chao, K.~R.~Parameswaran,
  M.~M.~Fejer, and I.~Brener, Opt.\ Lett.\ {\bf 24}, 1157 (1999).

\bibitem{gisin-PPLN} S.~Tanzilli,
H.~De~Riedmatten, W.~Tittel, H.~Zbinden, P.~Baldi, M.~De~Micheli,
D.~B.~Ostrowsky, and N.~Gisin, Electron.\ Lett.\, {\bf 37,} 26
(2001).

\bibitem{Sasha-theory} M.~H.~Rubin, D.~N.~Klyshko, Y.~H.~Shih, and
  A.~V.~Sergienko, Phys.\ Rev.\ A {\bf 50,} 5122 (1994).

\bibitem{rubin} M.~H.~Rubin, Phys.\ Rev.\ A {\bf 54,} 5349 (1996).

\bibitem{CRC} D.~R.~Lide, Ed. {\it CRC Handbook of Chemistry
and Physics}, 74th Ed. (CRC Press, Boca Raton, Fla., 1994), p.
{\bf 10}-304.

\end{references}
\end{document}